\documentclass[10pt,english,british,tightenlines,eqsecnum,
floats,aps,amsmath,amssymb,nofootinbib,superscriptaddress,prd,showpacs,showkeys]
              {revtex4}
\usepackage[latin9]{inputenc}
\usepackage{color}
\usepackage{babel}
\usepackage{amsmath}
\usepackage{amssymb}
\usepackage{graphicx}
\usepackage{esint}
\usepackage[unicode=true,pdfusetitle,
 bookmarks=true,bookmarksnumbered=false,bookmarksopen=false,
 breaklinks=false,pdfborder={0 0 1},backref=false,colorlinks=false]
 {hyperref}

\makeatletter
\@ifundefined{textcolor}{}
{%
 \definecolor{BLACK}{gray}{0}
 \definecolor{WHITE}{gray}{1}
 \definecolor{RED}{rgb}{1,0,0}
 \definecolor{GREEN}{rgb}{0,1,0}
 \definecolor{BLUE}{rgb}{0,0,1}
 \definecolor{CYAN}{cmyk}{1,0,0,0}
 \definecolor{MAGENTA}{cmyk}{0,1,0,0}
 \definecolor{YELLOW}{cmyk}{0,0,1,0}
}

\@ifundefined{date}{}{\date{02-July-2016}}
\usepackage{babel}

\usepackage{euscript}\usepackage{epsfig}

\usepackage{amsfonts}

\usepackage{fancyhdr}

\makeatother

\begin{document}

\title{Exact Cosmological Solutions in Modified Brans--Dicke Theory}

\author{S. M. M. Rasouli}

\email{mrasouli@ubi.pt}

\affiliation{Departamento de F\'{i}sica, Universidade da Beira Interior, Rua Marqu\^{e}s d'Avila
e Bolama, 6200 Covilh\~{a}, Portugal}

\affiliation{Centro de Matem\'{a}tica e Aplica\c{c}\~{o}es (CMA - UBI),
Universidade da Beira Interior, Rua Marqu\^{e}s d'Avila
e Bolama, 6200 Covilh\~{a}, Portugal}

\author{Paulo Vargas Moniz}

\email{pmoniz@ubi.pt}

\affiliation{Departamento de F\'{i}sica, Universidade da Beira Interior, Rua Marqu\^{e}s d'Avila
e Bolama, 6200 Covilh\~{a}, Portugal}

\affiliation{Centro de Matem\'{a}tica e Aplica\c{c}\~{o}es (CMA - UBI),
Universidade da Beira Interior, Rua Marqu\^{e}s d'Avila
e Bolama, 6200 Covilh\~{a}, Portugal}

\begin{abstract}
In this paper,
we obtain exact cosmological vacuum solutions for
an extended FLRW homogenous and isotropic Brans-Dicke (BD) universe in five
dimensions for all values of the curvature index.
Then, by employing the equations associated to a modified Brans-Dicke
theory (MBDT)~\cite{RFM14}, we construct the physics on a four-dimensional hypersurface.
We show that the induced matter obeys the equation of state of a fluid of a barotropic type.
We discuss the properties of such an induced matter for some values of the equation of state
parameter and analyze in detail their corresponding solutions.
To illustrate the cosmological behaviors of the solutions, we contrast our solutions with those
present the standard Brans-Dicke theory. We retrieve that, in MBDT scenario, it is impossible to find a
physically acceptable solution associated to the negative curvature for both
the dust-dominated and radiation-dominated universes. However, for a spatially flat and closed universes,
we argue that our obtained solutions are more
general than those associated to the standard BD theory and, moreover, they contain a few
 classes of solutions which have no analog in the BD cosmology.
  For those particular cases, we further compare the results with those
 extracted in the context of the induced matter theory (IMT) and general relativity (GR).
 Furthermore, we discuss in detail the time behaviors of the cosmological quantities and compare them
  with recent observational data.
 We find a favorable range for the deceleration parameter associated to a matter-dominated
 spatially flat universe (for the late times) which is compatible with recent corresponding observational results.
 \end{abstract}

\medskip



\maketitle

\section{Introduction}
\label{int} \indent
The scalar-tensor theories (for a complete review for applications in
cosmology, see, e.g., \cite{Faraoni.book}) have been proposed based on the main idea
 which asserts that the gravitational coupling is time-dependent, an idea
 related to the large number hypothesis by Dirac~\cite{D37,D38}.
 He suggested that the gravitational constant $G$ decreases
 with the age of the universe, an assumption that can be
  in agreement with a few consideration of geological facts~\cite{J55,J66}.
 Then, the Dirac's idea has been developed by other physicists by constructing
 new versions of the scalar-tensor theories in which the
 gravitational constant has been replaced by a scalar field~\cite{J48,T48}, and consequently it must satisfy a
 generalized conservation law proposed in the theory.
 In 1961, the simplest version of the scalar tensor theories was proposed
 by Brans and Dicke primarily motivated by
 cosmology and the Mach's principle~\cite{BD61}.

 Using different methods, exact cosmological solutions have been
 obtained for the isotropic FLRW models, the anisotropic Bianchi
 types I-IX and the related Kantowski-Sachs models by assuming various kinds of
 the ordinary matter in the context of the BD theory (see,
 e.g.,~\cite{o'hanlon-tupper-72,DO71,DO72a,DO72b,LP-rev,M78,C83,CN91,BP97,qiang2005,qiang2009,Faraoni.book,RFK11,LF15} and references therein).

 Recently, it has been shown that by applying a specific reductional procedure
 for the conventional BD theory in $(D+1)$-dimensional
 space-time, a MBDT in $D$ dimensions is obtained~\cite{RFM14}.
 The MBDT has four sets of field equations, in which two sets
 of them, regardless of the geometrical origin of the matter and scalar
 potential, are mathematically similar to those derived from the BD standard action with a scalar potential.
 One set of the mentioned MBDT field equations is the extended
 version of the conservation equation introduced in induced
 matter theory (IMT), see e.g.,~\cite{PW92,stm99,DRJ09,RJ10}. Finally, the forth equation retrieved in
 MBDT has no analog in the conventional BD theory as well as in scalar tensor theories.

 To gain an insight into the physical features of the MBDT and
 comparing the properties of its geometrically induced matter and scalar potential with
 the ordinary ones assumed in the context of the BD
 theory as well as with observational data, several investigations have been presented
 by considering both the spatially flat FLRW universe (in four~\cite{Ponce1,Ponce2} and arbitrary~\cite{RFM14}
 dimensions) and the Bianchi type I cosmology~\cite{RFS11}.
 We should note that, up to our knowledge, the solutions of the non-flat FLRW space-time have
 not been obtained in either the higher-dimensional BD theory or the MBDT scenario.
 Moreover, the herein solutions associated to flat
 space are wider than the ones obtained in~\cite{RFM14}.
 It is important to note that the induced scalar
 potential and matter in the MBDT lead us to an accelerating universe in late
 times (as it will be seen in this work) and we do not require to add a
 scalar potential by hand such as some phenomenological
 scenarios in the context of the DB theory~\cite{BP01,SS01}.

 The aim of this paper will be to employ the MBDT to reduce the five-dimensional
 FLRW solutions (for all values of the curvature) on a suitably projected four-dimensional
 space time. We will show that the induced matter obeys the barotropic
 equation of state and consequently this geometrically originated matter can be examined for
 particular choices of well-know matter types in the universe.
 Subsequently, the properties of obtained extended solutions will be analyzed and then
 they will be compared with the ones obtained from the four dimensional BD
  theory (with or without a scalar potential) as well as with observational data.
  We will further show that all the
 solutions can be considered as the generalized versions of
 the well-known solutions of the BD
 theory (e.g., the Dehnen-Obregon, Nariai, O'Hanlon-Tupper solutions) in four dimensional space-time.

 In this work, the expressions for physical quantities will be obtained neither by using
 the conformal time nor by defining other
 variables to explain the behavior of the scale factor and scalar field, but
 instead, we will discuss the behaviors from their corresponding direct physical expressions.

  Our work is organized as follows. In section~\ref{Set up}, we review
  the MBDT set up.
  In section~\ref{5d-cosmology}, by assuming a
  five-dimensional FLRW universe (with all values of the curvature), we solve the field
  equations associated to the standard BD theory in vacuum.
  In section~\ref{Reduced Cosmology}, by means of the MBDT setting, we construct the physics on
  a four-dimensional hypersurface. We show that the induced
  matter obeys the barotropic equation of state and consequently it
  can play the role of either the ordinary matter or dark matter-dark energy in the universe.
  By assuming a few known values for the equation of state parameter,
  we discuss the properties of the solutions and compare them with those
  obtained in the context of the BD theory as well as with observational data.
  Finally, in section~\ref{concl.}, we review the main results and add some new discussions.

\section{Modified Brans--Dicke Theory in Four Dimensions}
\label{Set up}
\indent
Let us in this section briefly review the MBDT set up~\cite{RFM14,Ponce2}.

In analogy to the four-dimensional standard
BD theory, the action associated to a five-dimensional BD theory, in the Jordan
frame, can be written as
\begin{equation}\label{(D+1)-action}
{\cal S}^{(5)}=\int d^{5}x \sqrt{\Bigl|{}{\cal
G}^{(5)}\Bigr|} \,\left[\phi
R^{(5)}-\frac{\omega}{\phi}\, {\cal
G}^{ab}\,(\nabla_a\phi)(\nabla_b\phi)+16\pi\,
L\!^{(5)}_{_{\rm matt}}\right],
\end{equation}
where the Latin indices run from zero to four, ${\cal G}^{(5)}$ is the determinant of the
metric ${\cal G}_{ab}$ associated to a five-dimensional
space-time, $R^{(5)}$ is the curvature scalar and $\nabla_a$
stands for the covariant derivative in the five-dimensional
space-time. Moreover, $\phi$ is the
BD scalar field, $\omega$ is the adjustable
dimensionless BD coupling
parameter and we have chosen $c=1$.
In addition, $L^{(5)}_{\rm matt}$ is the Lagrangian density associated to the
 ordinary matter which is minimally coupled to the BD scalar field.

The field equations, derived from the action (\ref{(D+1)-action}), are given by
\begin{equation}\label{(D+1)-equation-1}
G^{(5)}_{ab}=\frac{8\pi}{\phi}\,T^{(5)}_{ab}+\frac{\omega}{\phi^{2}}
\left[(\nabla_a\phi)(\nabla_b\phi)-\frac{1}{2}{\cal G}_{ab}(\nabla^c\phi)(\nabla_c\phi)\right]
+\frac{1}{\phi}\Big(\nabla_a\nabla_b\phi-{\cal G}_{ab}\nabla^2\phi\Big)
\end{equation}
and
\begin{equation}\label{(D+1)-equation-4}
\nabla^2\phi=\frac{8\pi T^{(5)}}{3\omega+4},
\end{equation}
where $\nabla^2\equiv\nabla_a\nabla^a$ and $T^{(5)}_{ab}$ is
the energy--momentum tensor (EMT) of the ordinary matter fields in a
five-dimensional space--time and $T^{(5)}={\cal G}^{ab}T^{(5)}_{ab}$.

By means of the reduction procedure for the BD theory,
the field equations associated to the four-dimensional MBDT can be obtained on a hypersurface.
In these field equations, geometrically induced terms will play
the role of the ordinary matter sources and scalar potential in an
extended version of the standard BD theory\footnote{In the Lagrangian density associated to
the original BD theory, there is no scalar
potential.}. Let us be more precise.
By applying the well-known line element\footnote{The Greek indices run
from zero to $3$ and $l$ stands for the
non--compact coordinate associated to fifth dimension. The indicator
$\epsilon=\pm1$ is chosen such that the extra dimension to be
either time--like or space--like, and $\psi=\psi(x^{\mu} ,l)$ is another
scalar field.}~\cite{stm99}
\begin{equation}\label{global-metric}
dS^{2}={\cal G}_{ab}(x^c)dx^{a}dx^{b}=
g_{\mu\nu}(x^\alpha,l)dx^{\mu}dx^{\nu}+
\epsilon\psi^2\left(x^\alpha,l\right)dl^{2},
\end{equation}
the BD field equations~(\ref{(D+1)-equation-1}) and
(\ref{(D+1)-equation-4}) are induced on the
hypersurface $\Sigma_0 (l=l_0={\rm constant}$), which is orthogonal to
the unit vector $n^a=\delta^a_{_D}/\psi$ where $n_an^a=\epsilon$.
In this respect, the MBDT, which can be regarded as a generalized version of the IMT, BD theory or GR,
conveys four field equations. One pair of these equations reproduces conventional
four-dimensional BD field equations but with the specificity that an induced scalar potential is present, in which the
induced EMT and the mentioned scalar potential have a geometrical origin.
The other pair has no analog in the standard four-dimensional BD theory
(for a more detailed presentation of the MBDT, see \cite{RFM14}).
We will now explain how our equations for a four-dimensional universe
is obtained from a five-dimensional BD setting, on a four-dimensional
hypersurface, directly and explicitly computing all terms, by means of
dimensional reduction and projection, i.e., all elements having a direct
geometric origin, computed consistently and none taken as an ad-hoc assumption.
Namely, in the next section, we will use

\begin{eqnarray}\label{BD-Eq-DD}
G_{\mu\nu}^{(4)}\!\!&=&\!\!\frac{8\pi}{\phi}\,
\left(S_{\mu\nu}+T_{\mu\nu}^{^{[\rm BD]}}\right)+
\frac{\omega}{\phi^2}\left[({\cal D}_\mu\phi)({\cal D}_\nu\phi)-
\frac{1}{2}g_{\mu\nu}({\cal D}_\alpha\phi)({\cal
D}^\alpha\phi)\right] \cr
 &&+\frac{1}{\phi}\left[{\cal D}_\mu{\cal
D}_\nu\phi- g_{\mu\nu}{\cal
D}^2\phi\right]-g_{\mu\nu}\frac{V(\phi)}{2\phi} \cr
 & \equiv &
\frac{8\pi}{\phi} T_{\mu\nu}^{(4)[{\rm eff}]} +
\frac{\omega}{\phi^2}\left[({\cal D}_\mu\phi)({\cal D}_\nu\phi)-
\frac{1}{2}g_{\mu\nu}({\cal D}_\alpha\phi)({\cal
D}^\alpha\phi)\right] + \frac{1}{\phi}\left[{\cal D}_\mu{\cal
D}_\nu\phi- g_{\mu\nu}{\cal D}^2\phi\right]-g_{\mu\nu}\frac{V(\phi)}{2\phi},
\end{eqnarray}
where ${\cal D}_\alpha$ is the covariant derivative on the hypersurface and
\begin{eqnarray}\label{D2-phi}
{\cal D}^2\phi=\frac{8\pi}{2\omega+3}\left(S+T^{^{[\rm BD]}}\right)+
\frac{1}{2\omega+3}\left[\phi\frac{dV(\phi)}{d\phi}-2V(\phi)\right],
\end{eqnarray}
where ${\cal D}^2\equiv{\cal D}_\alpha{\cal D}^\alpha$ and
\begin{eqnarray}\label{S}
S_{\mu\nu}\equiv T_{\mu\nu}^{(5)}-
g_{\mu\nu}\left[\frac{(\omega+1)T^{(5)}}{3\omega+4}-
\frac{\epsilon\, T_{44}^{(5)}}{\psi^2}\right],
\end{eqnarray}
is the effective matter obtained from the five-dimensional EMT and
\begin{eqnarray}\label{matt.def}
T_{\mu\nu}^{^{[\rm BD]}}= T_{\mu\nu}^{^{[\rm IMT]}}+T_{\mu\nu}^{^{[\rm \phi]}}
+\frac{1}{16\pi}g_{\mu\nu}V(\phi),
\end{eqnarray}
is an induced EMT on a four-dimensional space-time in which
\begin{eqnarray}\label{IMTmatt.def}
\frac{8\pi}{\phi}T_{\mu\nu}^{^{[\rm IMT]}}&\!\!\!\equiv &\!\!
\frac{{\cal D}_\mu{\cal D}_\nu\psi}{\psi}
-\frac{\epsilon}{2\psi^{2}}\left(\frac{\psi'
g'_{\mu\nu}}{\psi}-g''_{\mu\nu}
+g^{\lambda\alpha}g'_{\mu\lambda}g'_{\nu\alpha}
-\frac{1}{2}g^{\alpha\beta}g'_{\alpha\beta}g'_{\mu\nu}\right)\cr
 \!\!\!&&\!\!-\frac{\epsilon g_{\mu\nu}}{8\psi^2}
\left[g'^{\alpha\beta}g'_{\alpha\beta}
+\left(g^{\alpha\beta}g'_{\alpha\beta}\right)^{2}\right],\\\nonumber
\\
\label{T-phi} \frac{8\pi}{\phi}T_{\mu\nu}^{^{[\rm
\phi]}}&\!\!\!\equiv &\!\!
\frac{\epsilon\phi'}{2\psi^2\phi}\left[g'_{\mu\nu}
+g_{\mu\nu}\left(\frac{\omega\phi'}{\phi}-g^{\alpha\beta}g'_{\alpha\beta}\right)\right],
\end{eqnarray}
where a prime denotes a derivative with respect to the fifth coordinate, $l$.
The quantity $V(\phi)$ is an induced scalar potential
  on the hypersurface which is obtained from
\begin{eqnarray}\label{v-def}
\phi \frac{dV(\phi)}{d\phi}\!\!&\equiv&\!\!-2(\omega+1)
\left[\frac{({\cal D}_\alpha\psi)({\cal D}^\alpha\phi)}{\psi}
+\frac{\epsilon}{\psi^2}\left(\phi''-
\frac{\psi'\phi'}{\psi}\right)\right]-\frac{2\epsilon\omega\phi'}{2\psi^2}
\left[\frac{\phi'}{\phi}+g^{\mu\nu}g'_{\mu\nu}\right]\\\nonumber
&&\!\!\!
+\frac{\epsilon\phi}{4\psi^2}
\Big[g'^{\alpha\beta}g'_{\alpha\beta}
+(g^{\alpha\beta}g'_{\alpha\beta})^2\Big]
+16\pi\left[\frac{(\omega+1)T^{(5)}}{3\omega+4}-\frac{\epsilon T_{44}^{(5)}}{\psi^2}\right].
\end{eqnarray}
The first term of the induced EMT, i.e. $T_{\mu\nu}^{^{[\rm
IMT]}}$, which bears a resemblance to the one introduced in IMT~\cite{PW92,stm99}, is
 the fifth part of the metric (\ref{global-metric}) which is
geometrically induced on the hypersurface.
Whereas the $\phi$-part, i.e., $T_{\mu\nu}^{^{[\rm \phi]}}$, is composed of the BD scalar
field and its derivatives with respect to the fifth coordinate, has no analogue in IMT.
Moreover, the induced scalar potential is provided completely from the
geometry of the fifth dimension rather than adding an extra term by
hand, such as the ones usually employed in phenomenological
applications of the scalar tensor theories in
astrophysics/cosmology~see, e.g., \cite{SS01} and references therein.

In section~\ref{Reduced Cosmology}, we will use these geometrical
intrinsic as well as elegant advantages of the MBDT to
retrieve BD cosmological features on the hypersurface for a homogenous
and isotropic Friedmann universe and compare them with the results of
conventional BD theory, other scalar-tensor theories and also with observational data.

\section{exact solutions of Brans--Dicke cosmology in a five-Dimensional space-time}
\label{5d-cosmology}
\indent
We start with a five-dimensional extended version
of the~Friedmann-Lema\^{\i}tre-Robertson-Walker~(FLRW)
universe in which there is no ordinary matter, i.e., $T_{ab}^{(5)}=0$.
We will find the exact solutions in
the five-dimensional space-time and then, in
section~\ref{Reduced Cosmology}, by applying the set up reviewed in
the previous section, we concentrate on the cosmological
solutions further MBDT projected on a four-dimensional hypersuface.

Let us to work with a metric whose components depend only on
comoving time. Thus, we consider the line element
\begin{equation}\label{DO-metric}
dS^{2}=-dt^{2}+a^{2}(t)\left[\frac{dr^2}
{1-\lambda r^2}+r^2\left(d\theta^2+sin^2\theta d\phi^2\right)\right]+\epsilon \psi^2(t)dl^{2},
\end{equation}
where, without loss of generality, we can set the spatial curvature constant
as $\lambda=-1, 0, 1$, which correspond to open, flat and closed
universes, respectively; $(t, r, \theta, \varphi)$ are
the coordinates associated to a four-dimensional space-time whose
spatial sections are spherical symmetric. The $a(t)$ and $\psi(t)$ are
cosmological scale factors.

Employing equations~(\ref{(D+1)-equation-1}) and (\ref{(D+1)-equation-4}) with the line-element (\ref{DO-metric}) gives us the
dynamical field equations in a five-dimensional space-time, namely
\begin{eqnarray}\label{ohanlon-eq-1}
\ddot{\phi}+\dot{\phi}\left[\frac{\dot{3a}}{a}+\frac{\dot{\psi}}{\psi}\right]\!\!&=&\!\!0,\\
\label{ohanlon-eq-2}
\frac{\dot{a}}{a}\left(\frac{\dot{a}}{a}+\frac{\dot{\psi}}{\psi}\right)+\frac{\lambda}{a^2}
\!\!&=&\!\!-\frac{\dot{\phi}}{\phi}\left(\frac{\dot{a}}{a}
+\frac{\dot{\psi}}{3\psi}\right)+\frac{\omega}{6}\left(\frac{\dot{\phi}}{\phi}\right)^2,\\
\label{ohanlon-eq-3}
\frac{2\ddot{a}}{a}+\frac{\dot{a}}{a}
\left[\frac{\dot{a}}{a}+\frac{2\dot{\psi}}{\psi}\right]+\frac{\lambda}{a^2}
+\frac{\ddot{\psi}}{\psi}\!\!&=&\!\!-\frac{\dot{\phi}}{\phi}
\left(\frac{2\dot{a}}{a}+\frac{\omega}{2}\frac{\dot{\phi}}{\phi}
+\frac{\dot{\psi}}{\psi}\right)-\frac{\ddot{\phi}}{\phi},\\
\label{ohanlon-eq-4}
\frac{\ddot{a}}{a}+\left(\frac{\dot{a}}{a}\right)^2+\frac{\lambda}{a^2}
\!\!&=&\!\!-\frac{\dot{\phi}}{\phi}
\left(\frac{\dot{a}}{a}+\frac{\omega}{6}\frac{\dot{\phi}}{\phi}\right)-\frac{\ddot{\phi}}{3\phi}.
\end{eqnarray}
An overdot denotes the derivative
with respect to the cosmic time and we have assumed the BD
scalar field to depend only on the cosmic time. Equation~(\ref{ohanlon-eq-1}) has been derived
from~(\ref{(D+1)-equation-4}), and equations~(\ref{ohanlon-eq-2}),
(\ref{ohanlon-eq-3}) and (\ref{ohanlon-eq-4}) have been, respectively,
obtained by setting $a=0=b$, $a=b=1,2,3$ and $a=4=b$ into
Eq.~(\ref{(D+1)-equation-1}).

We should note that the coupled non-linear
equations~(\ref{ohanlon-eq-1})-(\ref{ohanlon-eq-4}) are not independent.
To solve these equations, let us consider a well-known assumption: we employ
 Dirac's hypothesis that states the gravitational
 constant and the scale factor of the universe
should be related to each other by a power law relation~\cite{DO71,DO72a,DO72b}.
As in the BD theory the scalar field is proportional to the
inverse of the gravitational constant~[see Eqs.~(\ref{G}) and~(\ref{G-PPN})], we can write
\begin{equation}\label{Dirac}
\phi(t)=C\left[a(t)\right]^n.
\end{equation}
In order to get consistent solutions, the above assumption
leads us to consider also a power law between $a(t)$ and $\psi(t)$, as
\begin{equation}\label{assumption-2}
\psi(t)=\alpha \left[a(t)\right]^{-\beta}.
\end{equation}
In relations (\ref{Dirac}) and (\ref{assumption-2}), $C>0$ (corresponding to attractive gravity) and $\alpha>0$ are
 constants which are not independent and they can be determined in an arbitrary
fixed time; $n$ and $\beta$ are
parameters which must satisfy the field equations\footnote{From current particle physics data, it can be assumed that for an extra
coordinate having lengthlike nature, it must be very small at
the present time~\cite{KS91,OW97}, i.e., it must contract with time. However, in this paper,
to get a comprehensive list of solutions, we will also investigate the cases in which $\beta<0$.}~(\ref{ohanlon-eq-1})-(\ref{ohanlon-eq-4}).

From (\ref{ohanlon-eq-1}), after a simple integration,
we get
\begin{equation}\label{gen-wave}
\dot{\phi}a^3\psi=A,
\end{equation}
where $A\neq0$ is a constant;
plugging $\phi$ and $\psi$ from (\ref{Dirac}) and (\ref{assumption-2}) into~(\ref{gen-wave}), we obtain
\begin{equation}\label{gen-scaleFactor}
a(t)=\left[\left(\frac{A}{C\alpha}\right)\frac{(3+n-\beta)t}{n}\right]^{1/(3+n-\beta)},
\end{equation}
where $3+n-\beta\neq0$ and we have assumed that at the initial time $t_i=0$, the scale factor vanishes.
By employing the power-law solutions (\ref{Dirac}) and (\ref{assumption-2}) in (\ref{ohanlon-eq-2}),
we obtain
\begin{equation}\label{Gen-eq}
\left(\frac{\dot{a}}{a}\right)^2\left[-\beta(1+\frac{n}{3})
+n\left(1-\frac{n\omega}{6}\right)+1\right]+\frac{\lambda}{a^2}=0.
\end{equation}

In what follows, we will investigate all the possible
solutions associated to the flat space ($\lambda=0$) and non-flat spaces ($\lambda^2=1$).

\subsection{Flat Space Solutions}
\label{F-5D}
For the flat space, by substituting $\lambda=0$
into (\ref{Gen-eq}), we obtain the following algebraic equation
\begin{equation}\label{Gen-eq-flat}
-\frac{\omega}{6}n^2+\left(1-\frac{\beta}{3}\right)n+(1-\beta)=0.
\end{equation}
Equation (\ref{Gen-eq-flat}) gives $n$ in terms of $\beta$ as
\begin{equation}\label{Gen-n-flat}
n=\frac{1}{\omega}\left[(3-\beta)\pm\sqrt{\beta^2-6\beta(1+\omega)+3(2\omega+3)}\right].
\end{equation}
Thus, the general solutions for the flat space are given by
relations~(\ref{Dirac}), (\ref{assumption-2})
and (\ref{gen-scaleFactor}) in which $n$ and $\beta$ are not independent
and they are related according to~(\ref{Gen-n-flat}). These solutions are classified into two sets:
 (i) the power-law solution when $n\neq\beta-3$, in which $n$ and $\beta$ are not independent and they
are related to each other by (\ref{Gen-n-flat}); (ii) the exponential solution that is obtained when $n=\beta-3$.
 To retrieve the second class of solutions, we have to start from the field
 equations~(\ref{ohanlon-eq-1})-(\ref{ohanlon-eq-4}). It is straightforward to show that
 \begin{equation}\label{Gen-eq-flat-exp}
 a(t)=a_ie^{\frac{At}{C\alpha(\beta-3)}},\hspace{15mm} \phi(t)=Ca_ie^{\frac{At}{C\alpha}},
 \hspace{15mm} \psi(t)=\alpha a_i^{-\beta}e^{\frac{-A\beta t}{C\alpha(\beta-3)}},
\end{equation}
where $a_i$ is an integration constant. For this particular
case, from (\ref{Gen-eq-flat}), $\beta$ in terms of the BD coupling parameter is given by
\begin{equation}\label{Beta-flat-exp}
 \beta=\frac{3(1+\omega)\pm\sqrt{-3(5+4\omega)}}{2+\omega}.
 \end{equation}
where, when $\omega\rightarrow-2$, for the upper case (plus sign) $\beta$ goes to unity while
for the lower sign, it goes to minus infinity. Namely, for the
former we get a contracting fifth dimension with time, while for the latter the fifth dimension increases by time.\\
\subsection{Non-Flat Space Solutions}
For $\lambda^2=1$, by employing (\ref{gen-scaleFactor}) into (\ref{Gen-eq}) we get an algebraic
relation which is satisfied for all values of $t$ provided that
\begin{equation}\label{Gen-n}
n=\beta-2,
\end{equation}
Thus, from (\ref{gen-scaleFactor}) the general solutions
associated to the non-flat space can be briefly summarized as
\begin{eqnarray}\nonumber
a(t)&=&\frac{1}{\beta-2}\left(\frac{A}{C\alpha}\right)t,
\hspace{3mm}\phi(t)=C\left[a(t)\right]^{\beta-2}=C\left[\frac{1}{\beta-2}\left(\frac{A}{C\alpha}\right)t\right]^{\beta-2},\hspace{3mm}{\rm and}\\
\hspace{3mm}\psi(t)&=&\alpha[a(t)]^{-\beta}=\alpha\left[\frac{1}{\beta-2}\left(\frac{A}{C\alpha}\right)t\right]^{-\beta}.\label{gen-solution}
\end{eqnarray}
where
\begin{equation}\label{Gen-AC-alpha}
\left(\frac{A}{C\alpha}\right)^2=\frac{6\lambda(\beta-2)^2}
{\omega(\beta-2)^2+2\beta(\beta-2)+6}.
\end{equation}
Besides, by employing relations~(\ref{ohanlon-eq-4}),
(\ref{Gen-n}), (\ref{gen-solution}) and (\ref{Gen-AC-alpha}),
it is easy to show that
\begin{eqnarray}\label{gen-omega}
\omega=-\frac{2(\beta^2-2\beta+3)}{(\beta-2)^2},
\end{eqnarray}
where in the particular case $\beta=2$, $|\omega|\rightarrow\infty$ and
consequently the BD scalar field takes a constant
value. In this case, we cannot
find a finite value for the scale factor $a(t)$. Namely, in this case,
when $|\omega|\rightarrow\infty$ there is no vacuum solution. As when the BD coupling parameter
goes to infinity, the BD theory may reduce to GR, thus the above result
is in agreement with the fact that there is no vacuum solutions
associated to the FLRW universe in GR (without a cosmological constant term) for $\lambda^2=1$.

In the special case where the scalar factor of the fifth dimension is a
constant, i.e., $\psi={\rm constant}$, by setting $\beta=0$, equation (\ref{Gen-AC-alpha}) gives
\begin{equation}\label{special-AC-alpha}
\left(\frac{A}{C\alpha}\right)^2=\frac{4\lambda}
{\frac{2\omega}{3}+1}.
\end{equation}
Therefore, for the cases with the positive curvature ($\lambda=1$)
and negative curvature ($\lambda=-1$), we must choose
$\omega>-3/2$ and $\omega<-3/2$, respectively.
From (\ref{special-AC-alpha}), for both of these
cases, we get $\frac{A}{C\alpha}=\pm\frac{2}{\sqrt{|1+\frac{2\omega}{3}|}}$, and thus by
substituting this relation into~(\ref{gen-solution}), we obtain
\begin{eqnarray}\label{special-solution}
a(t)=\mp\frac{t}{\sqrt{|1+\frac{2\omega}{3}|}}, \hspace{10mm}{\rm and}\hspace{10mm}
\phi(t)=\frac{C|1+\frac{2\omega}{3}|}{t^2}.
\end{eqnarray}

\section{Effective Brans-Dicke cosmology on a four dimensional hypersurface}
\label{Reduced Cosmology}

In this section, we will employ the MBDT set up reviewed in section~\ref{Set up},
for the obtained solutions of the previous section to
construct the physics on a four-dimensional space-time.
Then, we evaluate the properties of the corresponding solutions and
compare them with those of the standard BD
theory.

The non-vanishing components of the induced matter
on the hypersuface can be obtained by substituting the
components of the metric~(\ref{DO-metric}) and the
isotropic BD scalar field in relation~(\ref{matt.def}), as
\begin{eqnarray}\label{t-00}
\frac{8\pi}{\phi}T^{0[{\rm BD}]}_{\,\,\,0}\!\!\!&=&\!\!\!
-\frac{\ddot{\psi}}{\psi}+\frac{V(\phi)}{2\phi},\\\nonumber
\\
\label{t-ii}
\frac{8\pi}{\phi}T^{i[{\rm BD}]}_{\,\,\,i}\!\!\!&=&\!\!\!
-\frac{\dot{a}\dot{\psi}}{a\psi}+\frac{V(\phi)}{2\phi},
\end{eqnarray}
where $i=1,2,3$ (with no sum) and the induced potential $V(\phi)$
will be derived using the differential equation~(\ref{v-def}).
From~(\ref{t-ii}), it is clear that the different components of $T^{i[{\rm BD}]}_{\,\,\,i}$ are equal, thus
the induced matter can be considered as a perfect fluid.

The induced scalar potential is evaluated on the four-dimensional hypersurface
by substituting the solutions~(\ref{Dirac}), (\ref{assumption-2}),
(\ref{gen-scaleFactor}) and (\ref{gen-solution}) into (\ref{v-def}).
Thus, we obtain\footnote{We should notice that some quantities appearing in different solutions have
different values. For instance, the quantity $\frac{A}{C\alpha}$, which is seen in the upper
and lower relations of the induced scalar potential~(\ref{Eq-pot}),
 has different values in terms of the corresponding parameters;
 such that this quantity for the non-flat space is related to the
 other quantities by (\ref{Gen-AC-alpha}); whereas we have not
 obtained any general relation for this quantity in the case of the flat space.}
\begin{equation}\label{Eq-pot}
\frac{dV}{d\phi}{\Biggr|}_{_{\Sigma_{o}}}=\left \{
 \begin{array}{c}
 \frac{-2C^{\frac{2}{n}(3+n-\beta)}}{n}\beta(1+\omega)\left(\frac{A}{C\alpha}\right)^2
 \phi^{-\frac{2}{n}(3+n-\beta)}\hspace{10mm} {\rm for}\hspace{5mm} \lambda=0,
 \\
 \\
 -2C^{\frac{2}{\beta-2}}
(1+\omega)\left(\frac{\beta}{\beta-4}\right)\left(\frac{A}{C\alpha}\right)^2\phi^{\frac{-2}{\beta-2}}.
 \hspace{25mm} {\rm for}\hspace{5mm} \lambda=-1,+1.\\
 \end{array}\right.
\end{equation}
For a non-flat space, integrating the lower differential
equation of~(\ref{Eq-pot}), by assuming that the constants of integration are zero, we get
\begin{equation}\label{pot.gen}
V(\phi)=\left \{
 \begin{array}{c}
 \frac{-48\lambda C(1+\omega)}{2\omega+11}{\rm ln}\phi
 \hspace{15mm} {\rm for}\hspace{15mm} \beta=4\\
 \\
 -2C^{\frac{2}{\beta-2}}
(1+\omega)\left(\frac{\beta}{\beta-4}\right)\left(\frac{A}{C\alpha}\right)^2
\phi^{\frac{\beta-4}{\beta-2}} \hspace{15mm} {\rm for}\hspace{5mm} \beta\neq2,4,\\
 \end{array}\right.
\end{equation}
where $\frac{A}{C\alpha}$, according to relation (\ref{Gen-AC-alpha}), is a
function of the parameters $\beta$, $\omega$ and the curvature constant $\lambda$.

From the upper equation of~(\ref{Eq-pot}), the power-law scalar potential
associated to the flat space is given by
\begin{equation}\label{pot.gen-flat}
V(\phi)= \frac{2\beta(1+\omega)C^{\frac{2}{n}(3+n-\beta)}}{n-2\beta+6}\left(\frac{A}{C\alpha}\right)^2
 \phi^{\frac{-n+2\beta-6}{n}},\hspace{15mm} n\neq2(3-\beta),
 \end{equation}
where we have set the integration constant equal to zero.

The power-law scalar potential\footnote{As the logarithmic
effective potential leads to inconsistency for applying the energy conditions, we opt to
abandon the solutions associated to this case of the induced scalar potential.} [i.e., the lower relation in (\ref{pot.gen})
for non-flat space and (\ref{pot.gen-flat}) for flat space,
whose solutions, in what follows, will be discussed in this paper], vanishes in
the particular cases where either\footnote{In the special case where
$\omega=-1$, the BD theory corresponds to the low energy
limit of the bosonic string theory. Moreover, in this case, scalar-tensor
theories contain certain similarities with supergravity
and string theory~\cite{M86,Faraoni.book}.}$\omega=-1$ or $\beta=0$.

For the power-law scalar
potential, by employing~(\ref{gen-scaleFactor}) and~(\ref{gen-solution}) for flat
and non-flat spaces, we obtain the induced scalar potential versus cosmic time as
\begin{equation}\label{pot.gen-t}
V(\phi)=\left \{
 \begin{array}{c}
   2C\beta(n-2\beta+6)(1+\omega)\left(\frac{A}{C\alpha}\right)^{\frac{n}{3+n-\beta}}
 \left[\left(\frac{3+n-\beta}{n}\right)t\right]^{\frac{-n+2\beta-6}{n-\beta+3}}
  \hspace{15mm} {\rm for}\hspace{15mm}\lambda=0,\\
 \\
  -2C(1+\omega)\left(\frac{A}{C\alpha}\right)^{\beta-2}\left[\frac{\beta}{(\beta-4)(\beta-2)^{\beta-4}}\right]
t^{\beta-4} \hspace{25mm} {\rm for}\hspace{5mm} \lambda=-1,+1;\beta\neq2,4.
\end{array}\right.
\end{equation}

In order to write the components of the induced
matter (associated to the power-law cases) on the hypersurface,
we employ~(\ref{Dirac}), (\ref{assumption-2}), (\ref{gen-scaleFactor}), (\ref{Gen-AC-alpha}), (\ref{gen-solution}), (\ref{pot.gen-t}), and
relations~(\ref{t-00}) and~(\ref{t-ii}).  Thus, the energy density $\rho_{_{\rm BD}}\equiv -T^{0[{\rm BD}]}_{\,\,\,0}$
and the isotropic pressures $p_{_{\rm BD}}=p_i\equiv T^{i[{\rm BD}]}_{\,\,\,i}$ for the non-flat and flat spaces are given, respectively, by
\begin{equation}\label{ro-BD}
\rho_{_{\rm BD}}=\rho_{_{0}}t^{\beta-4}, \hspace{5mm}p_{_{\rm BD}}=p_{_{0}}t^{\beta-4},  \hspace{5mm}{\rm for} \hspace{5mm}\lambda=-1,+1,
\end{equation}
where
\begin{eqnarray}\label{rho-0}
\rho_{_{0}}&\equiv&\frac{C}{8\pi}\left(\frac{A}{C\alpha}\right)^{\beta-2}\beta
\left[\frac{(\beta+1)(\beta-4)+(1+\omega)(\beta-2)^2}{(\beta-4)(\beta-2)^{\beta-2}}\right],\\\nonumber\\
p_{_{0}}&\equiv&\frac{C}{8\pi}\left(\frac{A}{C\alpha}\right)^{\beta-2}\beta
\left[\frac{(\beta-4)-(1+\omega)(\beta-2)^2}{(\beta-4)(\beta-2)^{\beta-2}}\right]\label{rho-0},
\end{eqnarray}
and (for the flat space)
\begin{eqnarray}\label{ro-BD-flat}
\rho_{_{\rm BD}}&=&-\frac{C\beta}{8\pi}\left[\frac{A(n-\beta+3)}{Cn\alpha}\right]^{\frac{n}{n-\beta+3}}
\left[\frac{6(\beta-3)+n(n\omega+2\beta-9)}{(n-2\beta+6)(n-\beta+3)^2}\right]t^{\frac{n}{n-\beta+3}-2},\\\nonumber\\\nonumber\\\label{p-BD-flat}
p_{_{\rm BD}}&=&\frac{C\beta}{8\pi}\left[\frac{A(n-\beta+3)}{Cn\alpha}\right]^{\frac{n}{n-\beta+3}}
\left[\frac{(1+\omega)n^2+n-2\beta+6}{(n-2\beta+6)(n-\beta+3)^2}\right]t^{\frac{n}{n-\beta+3}-2}.
\end{eqnarray}
The above geometrical effective matter relations give
a barotropic equation of state associated to an induced perfect fluid for all the cases. Namely
\begin{eqnarray}\label{EOS.gen}
p_{_{\rm BD}}=W_{_{\rm BD}}\rho_{_{\rm BD}}, \hspace{10mm}
\end{eqnarray}
where
\begin{equation}\label{W-BD}
W_{_{\rm BD}}=\left \{
 \begin{array}{c}
 \frac{2(\beta-3)-n(n\omega+n+1)}{6(\beta-3)+n(n\omega+2\beta-9)}\hspace{10mm} {\rm for}\hspace{5mm} \lambda=0,
 \\
 \\
 \\
 \frac{(\beta-4)-(1+\omega)(\beta-2)^2}{(\beta+1)(\beta-4)+(1+\omega)(\beta-2)^2}\hspace{10mm} {\rm for}\hspace{5mm} \lambda=-1,+1\\
 \end{array}\right.
\end{equation}

By using the relations~(\ref{gen-scaleFactor}), (\ref{gen-omega}),~(\ref{EOS.gen}) and (\ref{W-BD}),
it is straightforward to show that the induced matter
(on the four-dimensional hypersurface) is conserved,
namely $\dot{\rho}_{_{\rm BD}}+3(\dot{a}/a)(\rho_{_{\rm BD}}+P_{_{\rm BD}})=0$.
This result is of relevance in the context of scalar
tensor theories, because it allows us to state that the BD
scalar field does couple minimally with the induced matter and thus, the herein
model does respect the Principle of the Equivalence.

As we would like to study, in particular, the behavior of the
gravitational coupling,
we review herewith general features associated to it.
As the field equations (\ref{BD-Eq-DD}) corresponds to the conventional BD field equations, the Newton's
gravitational constant (in a four-dimensional space-time) can be read~\cite{Faraoni.book}
\begin{equation}\label{G}
 G_{\rm N}(\phi)=\frac{1}{\phi}.
 \end{equation}
Moreover, we should note that the above definition has been used in cosmological
application, rather than in the analysis in the context of the Post-Newtonian formalism~\cite{W06}, where the relation
\begin{equation}\label{G-PPN}
 G_{\rm eff}=\frac{2(\omega+2)}{2\omega+3}\frac{1}{\phi},
 \end{equation}
has been defined for spherically symmetric solutions~\cite{N68,Faraoni.book, EP01,NP07} and
employed the Solar System experiments. However, we should note that the rate of variation of
$G_{\rm N}$ and $G_{\rm eff}$ have the same features.

In what follows, we will elaborate on the MBDT above retrieved for concrete values of
the equation of state parameter. After probing the properties of the
geometrical induced matter, we compare the results with those obtained
from the conventional BD theory and observational data.

\subsection{Vacuum Cosmologies}
One of the main questions in the context of scalar-tensor theories is whether non-trivial
vacuum (absence of ordinary matter) solutions
exist or not. Such solutions are very interesting because
 when $t$ asymptotically tends to zero, the fluid-filled FLRW solutions approach the
 vacuum solutions for wide ranges of the varying
 BD coupling parameter~\cite{B93}. After the seminal papers by Brans and Dicke, to
  our knowledge, this question has been investigated
by O'Hanlon and Tupper~\cite{o'hanlon-tupper-72}. They found the cosmological
non-static vacuum solutions for the BD theory for spatially
flat FLRW metric and addressed questions such as
the validity of the Mach's principle and Birkhoff theorem in the BD theory.
However, for the non-flat FLRW space, they only found solutions for spacial
 values of the BD coupling parameter, $\omega=-3/2,-4/3,0$.
 Subsequently, the whole class of solutions associated to the latter (non-flat) case
 have been obtained by Dehnen and Obreg\'{o}n by assuming a power law relation
 between the scale factor and the effective gravitational constant~\cite{DO72b}.

In what follows, we investigate the simple case in
which there are not any effects of the geometrical
induced matter and scalar potential on the MBDT
four-dimensional universe. In this case, we expect to get
solutions which would be similar to those obtained in the context of the conventional
 BD theory for a four-dimensional FLRW universe\rlap.\footnote{For convenience, from now on,
  we will investigate the solutions associated to the flat and non-flat spaces in separated parts.}

\subsubsection{\bf Flat space}
For the flat space ($\lambda=0$) and non-flat spaces ($\lambda^2=1$) the only manner
to get the vacuum cosmology on the hypersurface is setting $\psi=\alpha={\rm constant}$.
Therefore, for the first class of solutions (power-law solutions), by substituting $\beta=0$ into the solutions
associated to the flat space [i.e., relations (\ref{Dirac}), (\ref{assumption-2})
and (\ref{gen-scaleFactor}) in which $n$ and $\beta$
are related to each other by (\ref{Gen-n-flat})], after calculations, we get
\begin{eqnarray}\label{part-flat-solution}
a(t)&=&\left[\left(-\frac{A}{2C\alpha}\right)\left(1\mp3\sqrt{1+2\omega/3}
\right)t\right]^{r_{\pm}},\hspace{10mm} {\rm where}\hspace{10mm}r_{\pm}=\frac{(1+\omega)\mp\sqrt{1+2\omega/3}}{4+3\omega},\\\nonumber
\\
\label{par-phi-flat}
\phi(t)&=&C\left[a(t)\right]^{\frac{3}{\omega}(1\pm\sqrt{1+2\omega/3})},\\\nonumber
\end{eqnarray}
where we have assumed $1+2\omega/3\geq0$ and $\omega\neq-4/3$ (or $n\neq-3$).
For this case, from (\ref{Eq-pot}), without loss of generality, we can set the induced scalar
potential equal to zero, and thus, from using
relations~(\ref{t-00}) and (\ref{t-ii}), we find that the
induced matter also vanishes on the hypersurface.
The relations~(\ref{part-flat-solution}) and (\ref{par-phi-flat})
correspond to the well-known O'Hanlon and Tupper
solution~\cite{o'hanlon-tupper-72, RM14} which has been obtained for a
four-dimensional spatially flat FRW universe in the
standard BD theory where the ordinary matter is absent.
Let us present the above solutions in terms of Hubble parameter $H=\dot{a}/a$, and the age of the universe, $t_0$.
From~(\ref{part-flat-solution}), we get
\begin{equation}\label{uni.age}
 t_0=\frac{r_{\pm}}{H}.
 \end{equation}
Therefore, relations~(\ref{part-flat-solution}) and~(\ref{par-phi-flat}) can be written as
\begin{eqnarray}\label{part-flat-solution-t0}
a(t)&=&a_0\left(\frac{t}{t_0}\right)
^{r_{\pm}}
\hspace{5mm} {\rm with}\hspace{5mm} a_0=\left[\left(-\frac{A}{2CH\alpha}\right)
\left(1\mp\sqrt{1+2\omega/3}\right)\right]^{r_{\pm}},\hspace{5mm}
\\\nonumber
\\
\label{par-phi-flat-t0}
\phi(t)&=&\phi_0\left(\frac{t}{t_0}\right)^{m_{\pm}}\hspace{5mm} {\rm with}\hspace{5mm} \phi_0=\left[\left(-\frac{A}{2CH\alpha}\right)
\left(1\mp\sqrt{1+2\omega/3}\right)\right]^{m_{\pm}},\hspace{5mm}m_{\pm}=\frac{1\pm\sqrt{3(3+2\omega)}}{4+3\omega}\\\nonumber
\end{eqnarray}
where from the present values of the scale factor, Hubble constant and scalar field (i.e., $a_0$, $H_0$ and $\phi_0$, ), we can
determine the constant $A/C\alpha$.

\begin{figure}
\centering{}\includegraphics[width=3.3in]{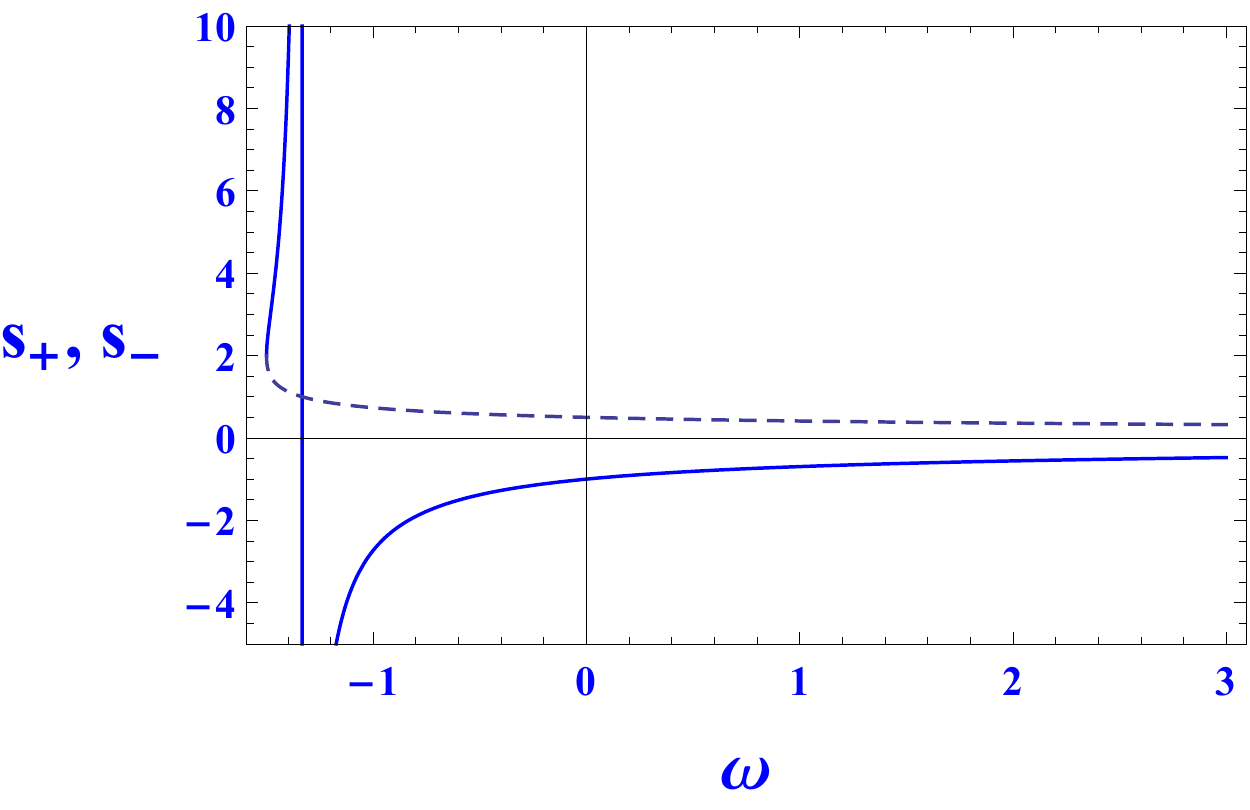}\hspace{4mm}
\includegraphics[width=3.3in]{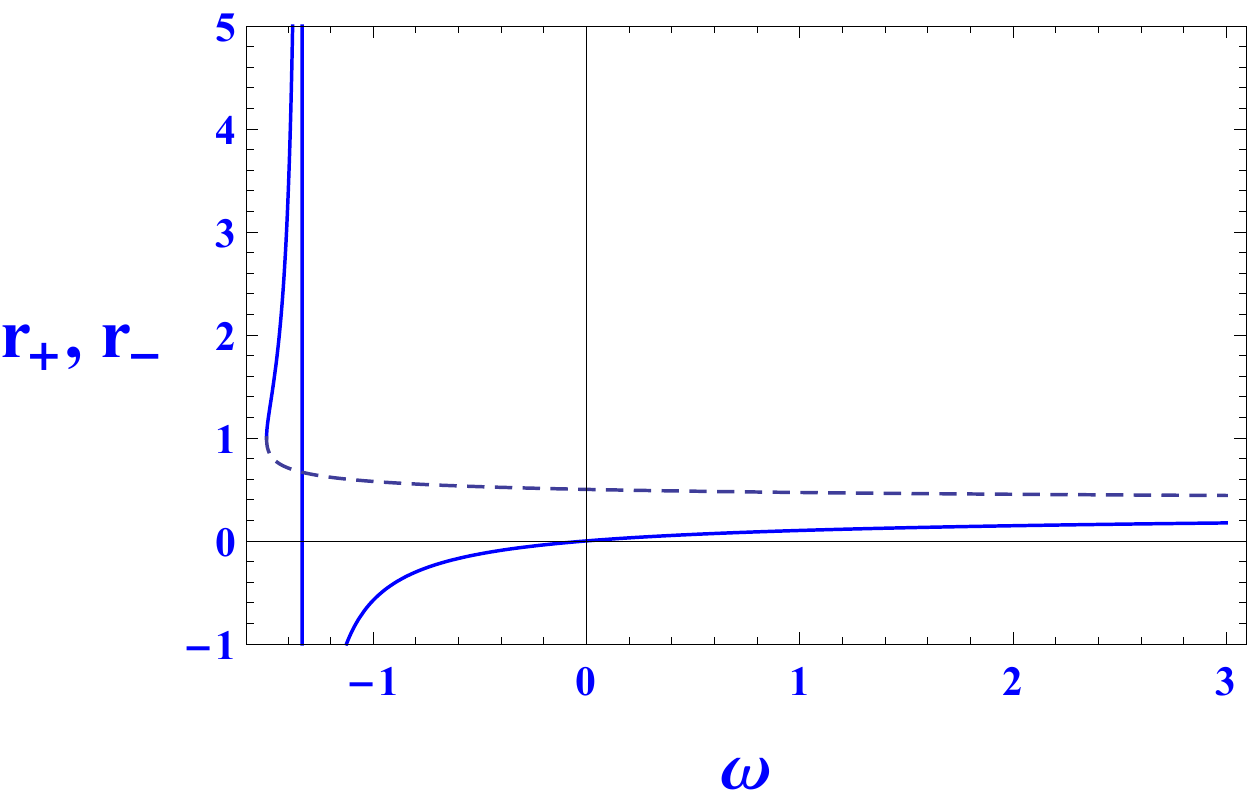}\hspace{4mm}
\caption{{\footnotesize The allowed ranges
for $s_+$, $r_+$ (the solid curves) and $s_{-}$, $r_-$ (the dashed curves) versus $\omega$
 for a power-law vacuum case which indicates
the time behavior of the gravitational constant and the scale factor.
The blue vertical line is associated to $\omega=-4/3$.}}
\foreignlanguage{english}{\label{s+-}}
\end{figure}

Let us now discuss the behavior of the gravitational
coupling, according to relation~(\ref{par-phi-flat-t0}), when $\omega>-3/2$  and $\phi_0>0$.
In\footnote{It will be of interest to discuss, for reason of convenience, the behavior of these and other
quantities by the least possible number of figures. However, unfortunately, as we will see, for some
cases, because of different ranges and scales used, it is not a feasible task.} Fig.~\ref{s+-} (the left panel), we have plotted the behavior of
$s_{\pm}\equiv-m_{\pm}$, the power of the cosmic time associated to the
gravitational coupling. As it is seen, for either the upper case (plus sign) (only when
the BD coupling parameter restricted to $-3/2<\omega<-4/3$) and
lower case (minus sign), the behavior of the
gravitational coupling is contrary to Dirac's hypothesis. Whereas, for the upper case, when $\omega>-4/3$ the gravitational
coupling decreases with cosmic time which is in accordance with Dirac's hypothesis.
Regarding the behavior of the scale factor, we see that only $\omega$ is
restricted as $-3/2<\omega<-4/3$, we get $r_{+}>1$, (see Fig. \ref{s+-}, the right panel),
 namely, which corresponds to an accelerating behavior of the scale factor for the upper case.
 The scale factor of the universe decelerates when either $\omega>-3/2$ (for the lower case) or $\omega>0$ (for the upper case).
From~(\ref{uni.age}) for the upper case, we see that when the BD coupling parameter
is larger than $-3/2$ and approaches to $-4/3$, the value of the age of the universe takes large values.
For the second class (exponential solution), by setting $\beta=0$
into relations (\ref{Gen-eq-flat-exp}) and (\ref{Beta-flat-exp}), we find
the solutions associated to $\omega=-4/3$ for the upper case (plus sign) as
\begin{equation}\label{flat-exp-zero.Beta}
 a(t)=a_ie^{-\frac{At}{3C\alpha}},\hspace{15mm} \phi(t)=Ca_i^{-3}e^{\frac{At}{C\alpha}}.
 \hspace{15mm} \psi(t)=\alpha={\rm constant},
\end{equation}
whereas that the lower case does not give acceptable solution.

 In other words, we have
\begin{equation}\label{flat-exp-zero.Beta-H}
 a(t)=a_ie^{Ht},\hspace{15mm} \phi(t)=Ca_i^{-3}e^{-3Ht};
 \hspace{10mm}{\rm with}\hspace{10mm} H=-\frac{A}{3C\alpha}={\rm constant},
\end{equation}
which indicates that the O'Hanlon and Tupper solution in the limiting
case $\omega=-4/3$ approaches to the de Sitter space, likewise
 the conventional BD theory, see, e.g., \cite{Faraoni.book}.
Similar to the first class of the flat space with $\beta=0$, for this
case also the components of the induced matter and the scalar potential vanish.
We should note that, similar to the standard BD theory,
 the solution~(\ref{flat-exp-zero.Beta-H}) is the unique de Sitter solution associated to the flat
 and vacuum space; and it is different from the one obtained in GR
 with a minimally coupled scalar field, in which that scalar
 field takes a constant value. By assuming
 that $\phi>0$, from~(\ref{flat-exp-zero.Beta-H}), we see that the BD
 scalar field decreases with cosmic time and thus the effective gravitational
 coupling~(\ref{G}) increases with time; such a behavior is contrary to Dirac's hypothesis.

\subsubsection{\bf Non-flat space}
It should be noted that there are no vacuum solutions associated to FLRW
space with $\lambda=\pm1$ in GR in the absence of the cosmological constant~\cite{LP83a}.
However, such solutions have been found in the BD theory~\cite{CE83,C83,LP83a}.
In what follows, we would like to investigate the corresponding solutions in MBDT.

In the case of the non-flat space where $\beta=0$, we get $\psi=\alpha={\rm constant}$ and the scale
factor and scalar field are given
by~(\ref{special-solution}).
For this case, from relations~(\ref{v-def}), (\ref{assumption-2}), (\ref{t-00}), and~(\ref{t-ii}),
we find that the induced matter and scalar potential vanish on the hypersurface.
 The cosmology associated to this case is the same as that obtained by Dehnen and Obreg\'{o}n in \cite{DO72b}.
 With $t_0=1/H$, the age of the universe at
present time\footnote{The age of the universe estimated by best
fit to the Planck 2013 data \cite{Planck2013} has been reported
$13.813\pm0.058$ billion years.}, the solutions~(\ref{special-solution}) can be given by
\begin{eqnarray}\label{special-solution-age}
a(t)=a_0\left(\frac{t}{t_0}\right), \hspace{10mm}{\rm and}\hspace{10mm}
\phi(t)=\phi_0\left(\frac{t_0}{t}\right)^2,
\end{eqnarray}
in which the present values associated to the scale factor and the BD scalar field are
\begin{eqnarray}\label{special-solution-values}
a_0=C\frac{1/H}{\sqrt{|1+\frac{2\omega}{3}|}}, \hspace{10mm}{\rm and}\hspace{10mm}
\phi_0=C\frac{|1+\frac{2\omega}{3}|}{(1/H)^2}.
\end{eqnarray}

 We should note that by determining $t_0$, $a_0$
 and $\phi_0$, the solutions associated to the
 closed and open universes cannot be distinguished
 if the BD coupling parameter restricted to $\omega>-3/2$ and $\omega<-3/2$, respectively.

\subsection{Dust Cosmologies}
Dust solutions associated to the FLRW universe
in the context of the BD theory (with vanishing scalar potential)
 have been investigated in \cite{BD61, DO71}. In \cite{M82,M01}, some
 properties of the mentioned solutions have been discussed for large values of $|\omega|$.
 Furthermore, there are a few publications in which the solutions have
 been obtained in terms of the conformal time and other variables
 different from the scale factor and the BD scalar field~\cite{MW95,TV96, O97,Faraoni.book}.
\subsubsection{\bf flat space}
For this case, by solving equations (\ref{W-BD}) (upper case) with $W_{_{\rm BD}}=0$ together
with the general equation (\ref{Gen-eq-flat}) associated to the flat
space, we obtain two classes of solutions as
\begin{eqnarray}\label{f-dust-n-beta}
n^{\pm}=\pm\frac{\sqrt{-3(1+\omega)}}{1+\omega},
\hspace{10mm} \beta^{\pm}=\frac{1}{2}\left(3\pm\frac{\sqrt{-3(1+\omega)}}{1+\omega}\right)
\end{eqnarray}
Thus, by substituting the above values for $\beta$ and $n$
into (\ref{Dirac}), (\ref{gen-scaleFactor}), first relation of
(\ref{pot.gen-t}) and (\ref{rho-0}) the solutions associated to the flat space are given by
\begin{eqnarray}\label{flat-dust-sol-a}
a^{\pm}&=&a_0^{\pm}t^{r^{\pm}}, \hspace{10mm}a_0^{\pm}=\left[\frac{A}{2C\alpha}\left(1-\sqrt{-3(1+\omega}\right)\right]^{r^{\pm}}\\
\label{flat-dust-sol-phi}
\phi^{\pm}&=&\phi_0^{\pm}t^{m^{\pm}},\hspace{10mm}\phi_0^{\pm}=C\left[\frac{A}{2C\alpha}\left(1-\sqrt{-3(1+\omega}\right)\right]^{m^{\pm}}\\
\label{flat-dust-sol-V}
V^{\pm}&=&V_0^{\pm}t^{\frac{2\left[-3(1+\omega)\pm\sqrt{-3(1+\omega)}\right]}{4+3\omega}}\\
\label{flat-dust-sol-rho}
\rho_{_{\rm BD}}^{\pm}&=&\rho_0^{\pm}t^{\frac{2\left[-3(1+\omega)\pm\sqrt{-3(1+\omega)}\right]}{4+3\omega}},
\end{eqnarray}
where
\begin{eqnarray}\label{m-dust}
r^{\pm}&=&\frac{2\left[(1+\omega)\pm\sqrt{-\frac{1+\omega}{3}}\right]}{4+3\omega},\\
m^{\pm}&=&n^{\pm}r^{\pm}=\frac{2\left[1\pm\sqrt{-3(1+\omega)}\right]}{4+3\omega},\\\nonumber\\
V_0^{\pm}&=&\frac{C}{3}\left[3(1+\omega)\pm\sqrt{-3(1+\omega)}\right]\left(\frac{A}{C\alpha}\right)
^{\frac{2\left[1\pm\sqrt{-3(1+\omega)}\right]}{4+3\omega}},\\\nonumber\\
\rho_0^{\pm}&=&\frac{-C\left[(5+4\omega)\pm\sqrt{-\frac{(1+\omega)}{3}}\right]}{4\pi(4+3\omega)}\left[-\frac{A}{2C\alpha}(1\pm\sqrt{-3(1+\omega)})\right]
^{\frac{2\left[1\pm\sqrt{-3(1+\omega)}\right]}{4+3\omega}},
\end{eqnarray}
where $\omega\neq-4/3$, $\omega<-1$ and the parameters
$r^{\pm}$ and $m^{\pm}$ are related to each other as $3r^{\pm}+m^{\pm}=2$.
  For this case, the scale factor of the fifth dimension is given by
\begin{eqnarray}\label{psi-dust}
\psi^{\pm}=\alpha\left[-\frac{A}{2C\alpha}\left(1\pm\sqrt{-3(1+\omega}\right)t\right]^{-1}.
\end{eqnarray}

\begin{figure}
\centering{}
\includegraphics[width=3.3in]{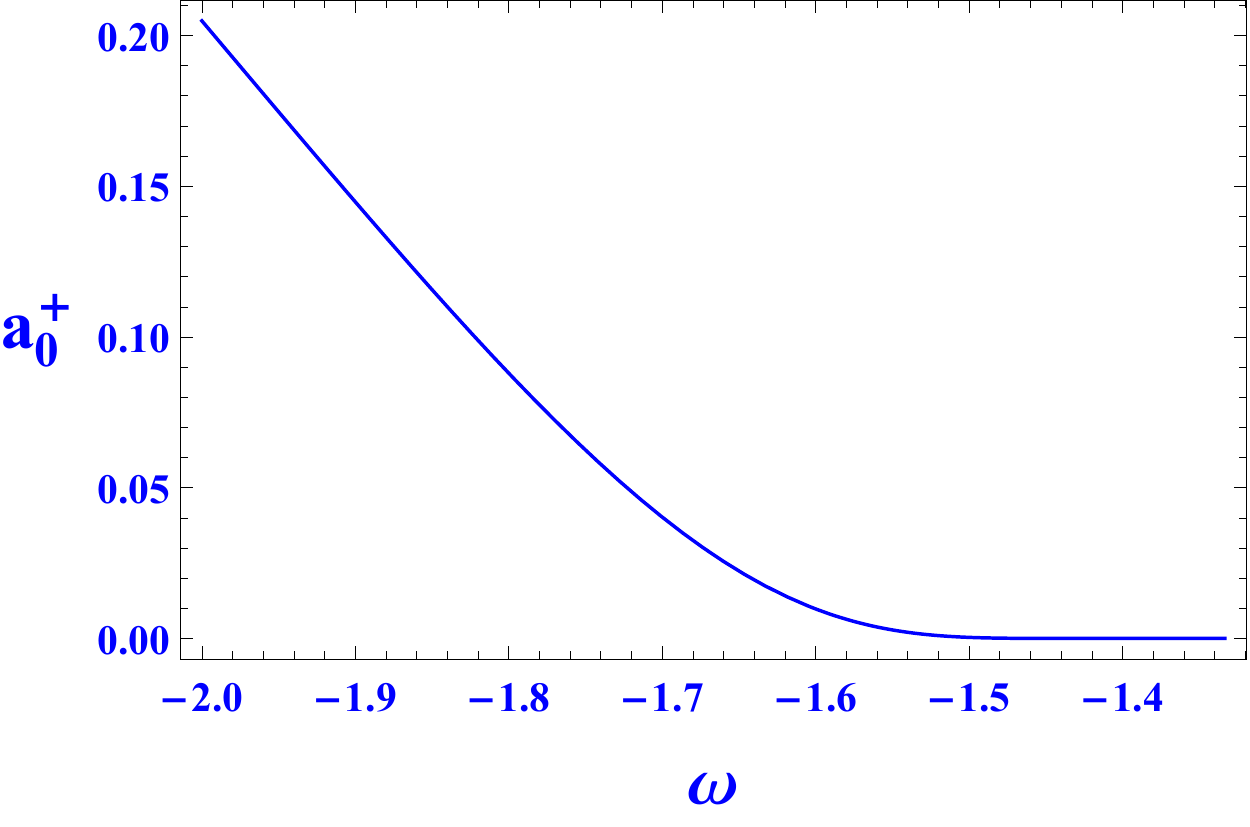}\hspace{4mm}
\includegraphics[width=3.3in]{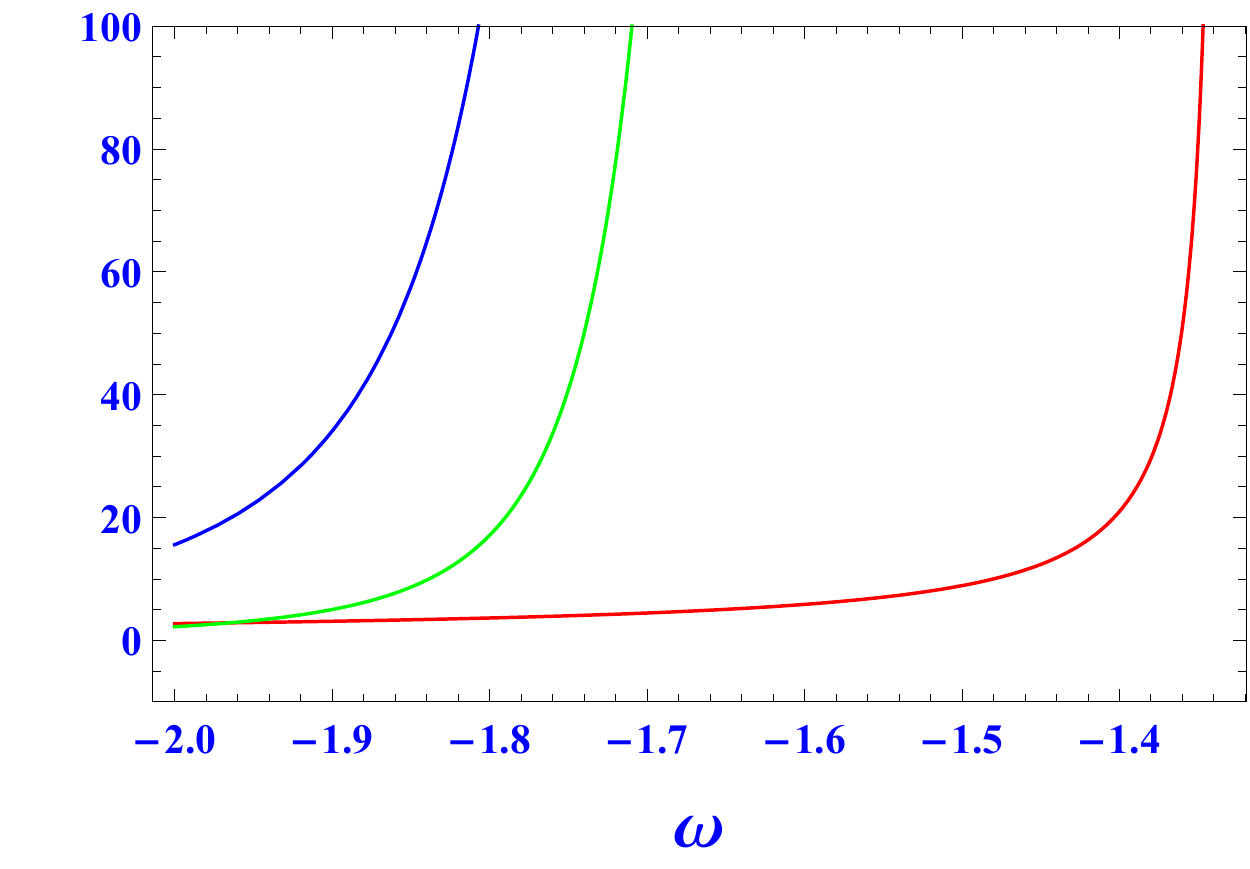}\hspace{4mm}
\caption{{\footnotesize The left panel shows the behavior of $a_0^+$ in the parameter space $(\omega, a_0^+)$.
In the right panel, blue, red and green curves, respectively, correspond to $\phi$,
$\psi$ and $\rho_{_{\rm BD}}$ versus $\omega$ for an arbitrary fixed time when $A<0$.
The mentioned quantities are associated to the spatially flat FLRW universe which is filled with
a dust fluid. We have set $A=-1$, $C=\alpha=1$. Note that when
$\omega$ tends to $-4/3$, $a_0^+$ takes very small values while $\rho_0^+$ takes very large values. }}
\foreignlanguage{english}{\label{flat-dust-q1}}
\end{figure}

\begin{figure}
\centering{}
\includegraphics[width=3.3in]{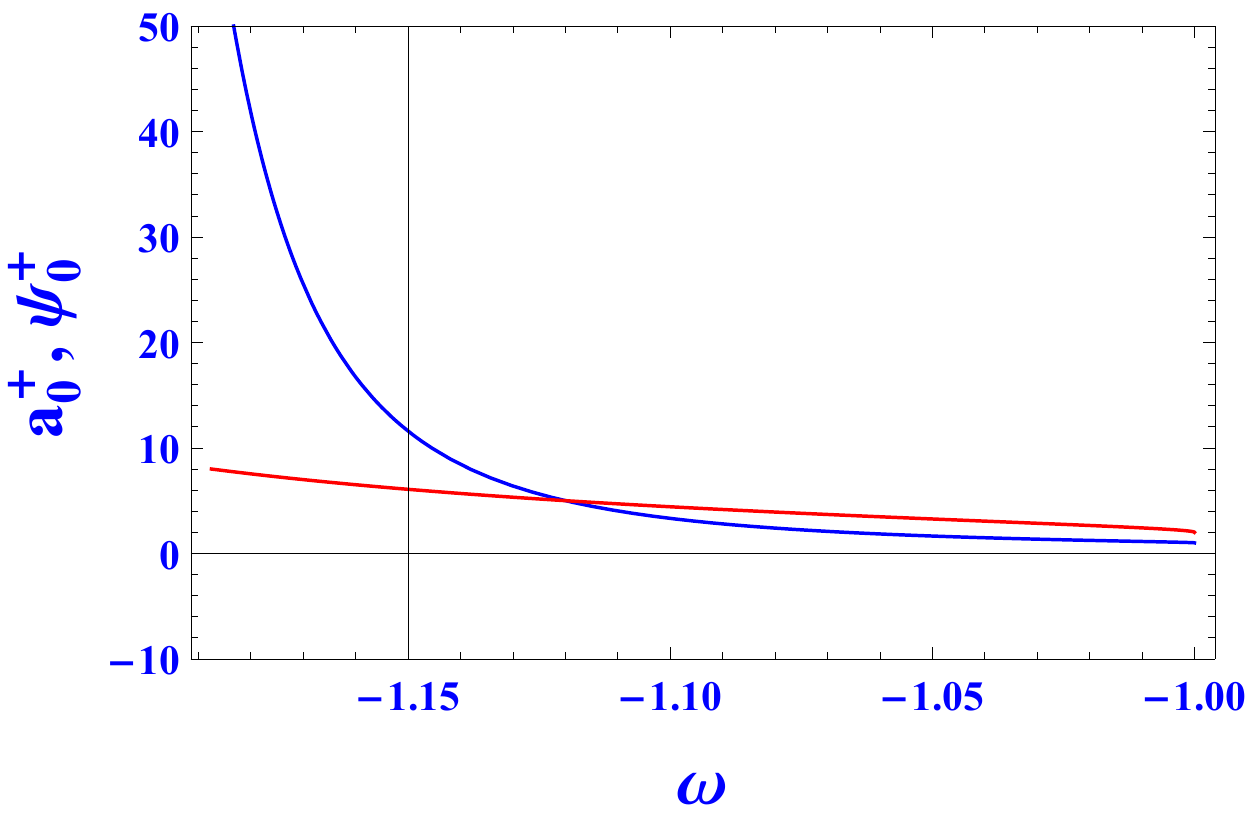}\hspace{4mm}
\includegraphics[width=3.3in]{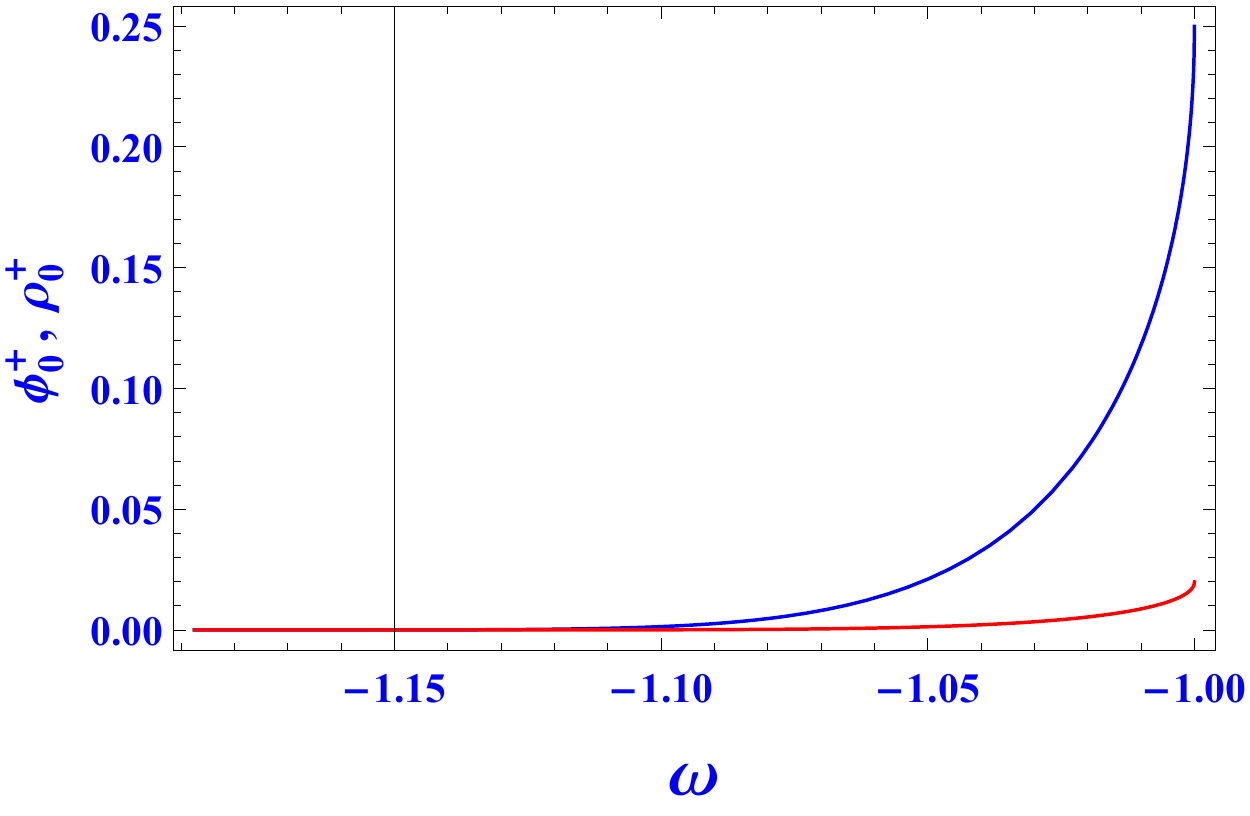}\hspace{4mm}
\includegraphics[width=3.3in]{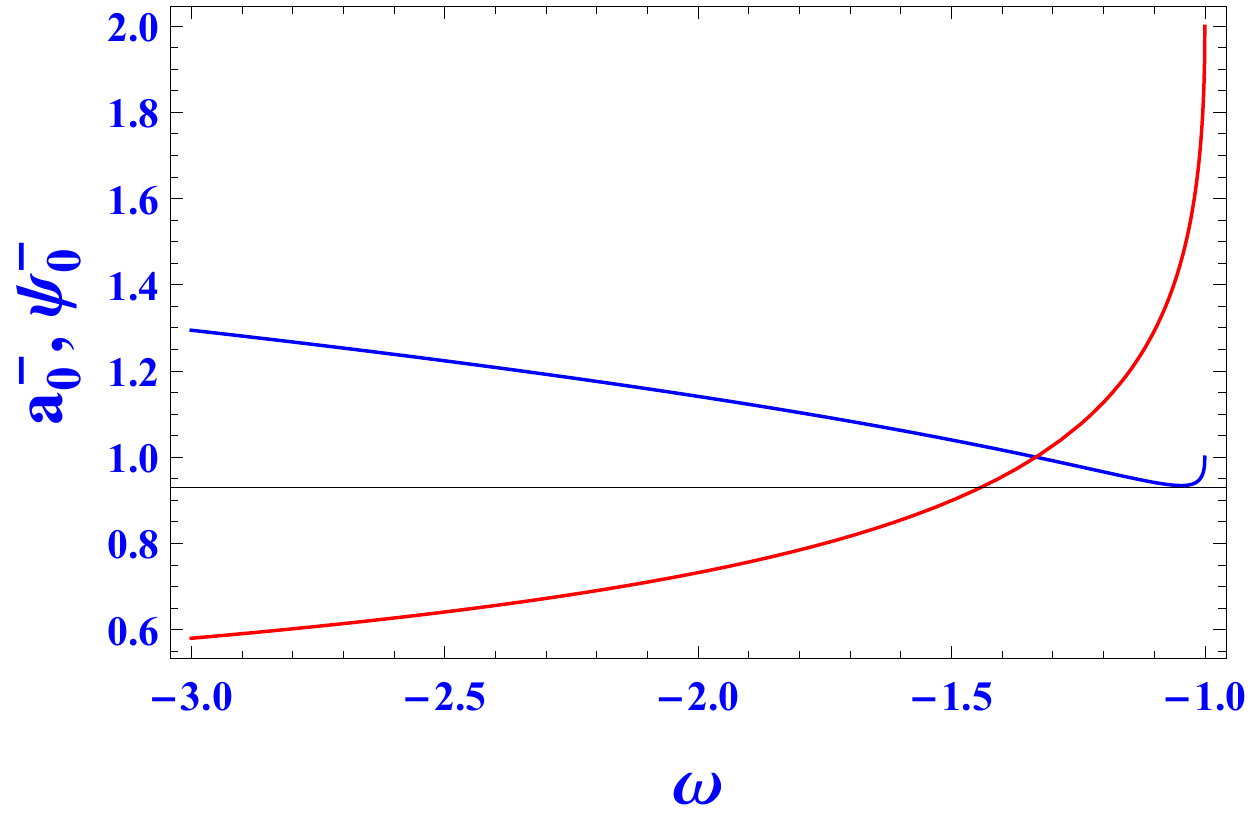}\hspace{4mm}
\includegraphics[width=3.3in]{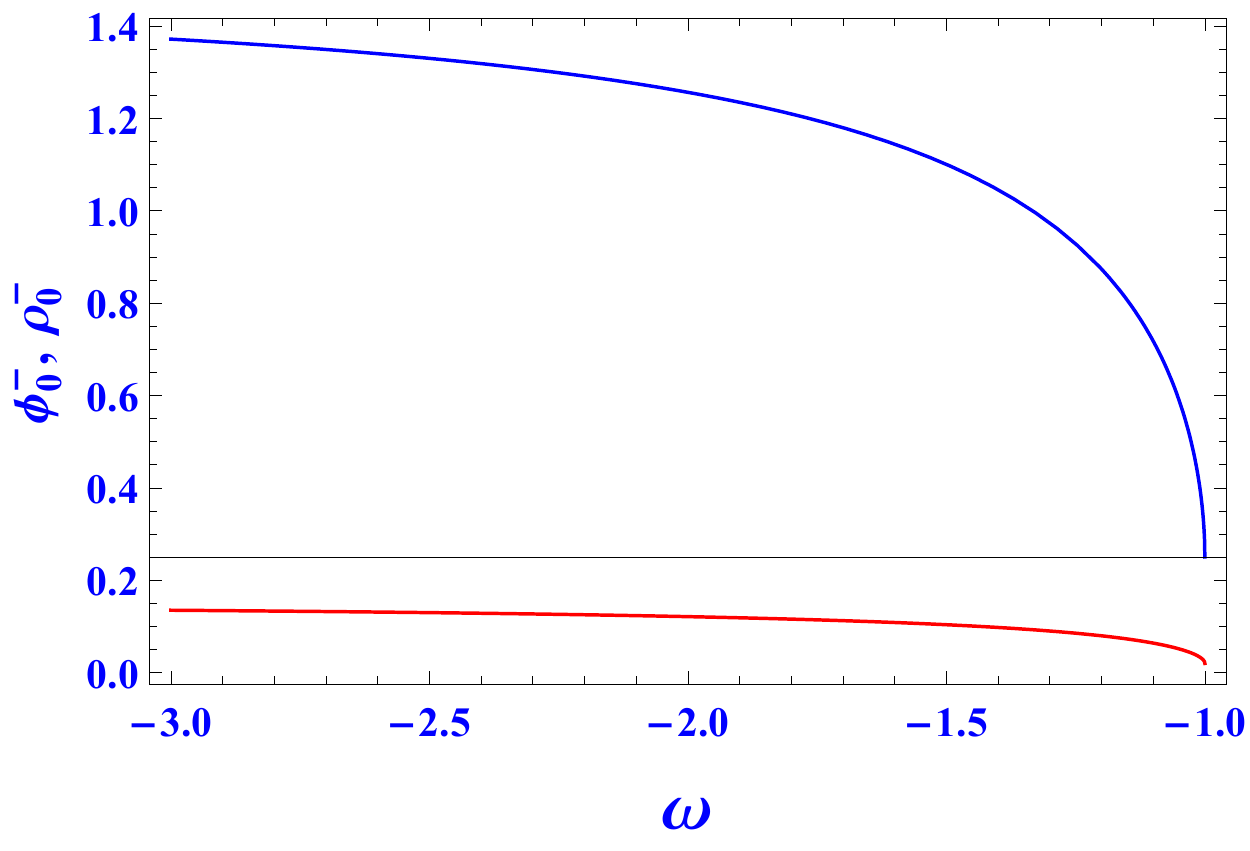}\hspace{4mm}
\caption{{\footnotesize The behavior of $\{a,\phi\}$ (the blue curves) and $\{\psi,\rho_{_{\rm BD}}\}$ (the red curves)
versus $\omega$ for an arbitrary fixed time when $A>0$. The mentioned quantities are
associated to the spatially flat FLRW universe which is filled with
a dust fluid. The upper and lower panels are associated to the upper and lower cases, respectively.
We have set $A$, $C$ and $\alpha$ equal to unity.
}}
\foreignlanguage{english}{\label{flat-dust-quantities}}
\end{figure}

As $a$, $\rho_{_{\rm BD}}$, $\phi$ and $\psi$ must take positive values for
an arbitrary fixed time, therefore, we can
find the allowed ranges for the BD coupling parameter for these solutions.
We should note that to have reasonable solutions, $C$ and $\alpha$ must
take positive values, but $A$ can take either positive or negative values,
 each of which gives two classes of solutions, one for the upper case and another for the lower case.
Let us categorize the resulted solutions in terms of the sign of $A$.
\begin{description}
  \item[{\bf Case I; $A<0$}:] For the lower case (minus sign), as the induced
  energy density and the scale factor do not take real values in
  an arbitrary fixed time, thus we abstain from considering them as acceptable solutions.
  However, for the upper case (plus sign), when the BD coupling parameter
  restricted to $\omega<-4/3$, we found that $a_0^+$, $\rho_0^+$, $\phi_0^+$ and $\psi_0^+$
  take positive real values (see Fig.\ref{flat-dust-q1}) and consequently this case can be a physical solution.
  In this setting, we see that for the allowed values of $\omega$, $r^{+}>1$, i.e., we get
  an accelerating universe. In addition, $m^+<0$ which means that the BD
  coupling parameter decreases with the cosmic time. Namely, the
  gravitational coupling increases with time which is not in agreement with Dirac's hypothesis.
  Moreover, the induced energy density and the fifth dimension decrease with time.
  \item[{\bf Case II; $A>0$:}]
For the lower case (minus sign) the quantities $a$, $\phi$, $\psi$
and $\rho_{_{\rm BD}}$ (in an arbitrary fixed time) take positive real values when the BD
coupling parameter is restricted to $\omega<-1$.
Whereas for the upper case (plus sign), we find a narrow range for the
BD coupling parameter which gives acceptable solutions.
Although in this case, $a$, $\phi$ and $\psi$ (for an arbitrary time) take positive real
values when $-4/3<\omega<-1$, but as $\rho$ is negative when $-4/3<\omega<-19/16$,
therefore the allowed range is $-19/16<\omega<-1$ in which the solution is acceptable.
In Fig.~\ref{flat-dust-quantities}, for the corresponding allowed ranges, we have shown the behavior of the
mentioned quantities versus $\omega$ for an arbitrary fixed time.

Let us summarize briefly the behavior of the quantities associated
to this case for the corresponding allowed ranges:
(i) we found that, for the lower case, the scale
factor of the universe decelerates with cosmic time while
we have a contracting scale factor for the upper case.
(ii) Both of $m^{+}$ and $m^-$ take positive values, namely, the BD scalar field
always increases with cosmic time and, consequently, the gravitational
coupling decreases with time which is in agreement with Dirac's hypothesis.
(iii) for the lower case, both the induced energy density
and the scale factor of the fifth dimension decreases with time.
However, for the upper case, the induced energy
density increases with time while the fifth dimension contracts with time.
\end{description}

\subsubsection{\bf non-flat space}
For this case, by considering the lower equation of (\ref{W-BD}) and solving $W_{_{\rm BD}}=0$, we get
two values for $\beta$ which both of them depend only on the BD coupling parameter as
\begin{eqnarray}\label{nf-Dust-beta}
\beta_i=\frac{(5+4\omega)+\chi_i}{2(1+\omega)},\hspace{10mm} i=1,2,\hspace{10mm}
\chi_1\equiv\sqrt{-(7+8\omega)},\hspace{10mm}\chi_2\equiv-\sqrt{-(7+8\omega)}
\end{eqnarray}
where $\omega\leq-7/8$ and $\omega\neq-1$. We should note that when $\omega$ tends to $-1$,
$\beta_1$ and $\beta_2$ goes to infinity and $4$, respectively.\\

By substituting $\beta_i$ from~(\ref{nf-Dust-beta}) into~(\ref{Gen-AC-alpha}), we get
\begin{equation}\label{X}
X_i\equiv\left[\left(\frac{A}{C\alpha}\right)^2\right]_i=
\frac{3\lambda\left[(3-8\omega)+7\chi_i\right]}
{(4+\omega)(11+8\omega)}.
\end{equation}

Replacing $\beta_i$ and $X_i$ from (\ref{nf-Dust-beta}) and (\ref{X}) into
(\ref{gen-solution}), the lower relation of (\ref{pot.gen-t}) and (\ref{ro-BD}), we get
four classes of mathematical solutions as\footnote{Indeed, we have two kinds of
solutions specified with $i=1,2$ (the first solution with $i=1$ and the
second solution with $i=2$), which they, in turn, have two cases specified as an
upper case (plus sign) and a lower case (minus sign).}
\begin{eqnarray}\nonumber
a_i^{\pm}&=&a_{i0}^{\pm}t,\hspace{23mm}\phi_i^{\pm}=\phi_{i0}^{\pm}t^{\frac{1+\chi_i}{2(1+\omega)}},\hspace{10mm} \psi_i^{\pm}=\psi_{i0}^{\pm}t^{-\frac{(5+4\omega)+\chi_i}{2(1+\omega)}},\nonumber\\\nonumber\\
 \rho_{_{\rm BDi}}^{\pm}&=&\rho_{i0}^{\pm}t^{-\frac{(3+4\omega)+\chi_i}{2(1+\omega)}},\hspace{7mm}
 V_i^{\pm}=V_{i0}^{\pm}t^{\frac{-(3+4\omega)+\chi_i}{2(1+\omega)}}\label{nf-dust-gen-1}
\end{eqnarray}
where
\begin{eqnarray}\nonumber
a_{i0}^{\pm}&=&\pm\frac{1}{4}\sqrt{3X_i}(1-\chi_i),\hspace{10mm}\phi_{i0}^{\pm}=C\left[a_{i0}^{\pm}\right]^{\frac{1+\chi_i}{2(1+\omega)}},\hspace{10mm}
 \psi_{i0}^{\pm}=\alpha\left[a_{i0}^{\pm}\right]^{-\frac{(5+4\omega)+\chi_i}{2(1+\omega)}},\\\nonumber\\\nonumber
 \rho_{i0}^{\pm}&=&-\frac{C}{16\pi}
 \left[\pm\sqrt{X_i}\right]^{\frac{1+\chi_i}{2(1+\omega)}}
 \left[\frac{1+\chi_i}{2(1+\omega)}\right]^{-\frac{(5+4\omega)+\chi_i}{2(1+\omega)}}
 \left\{\frac{4\omega^2[(29+8\omega)+\chi_i]+6\omega(23+\chi_i)+(53+\chi_i)}{(1+\omega)^4}\right\},\\\nonumber\\\label{nf-dust-gen-2}
  V_{i0}^{\pm}&=&-C\left[\pm\sqrt{X_i}\right]^{\frac{1+\chi_i}{2(1+\omega)}}
  \left[\frac{1+\chi_i}{2(1+\omega)}\right]^{-\frac{(5+4\omega)+
  \chi_i}{2(1+\omega)}}\left[\frac{-(11+20\omega+8\omega^2)+\chi_i}{(1+\omega)^3}\right].
\end{eqnarray}

\begin{figure}
\centering{}\includegraphics[width=3.3in]{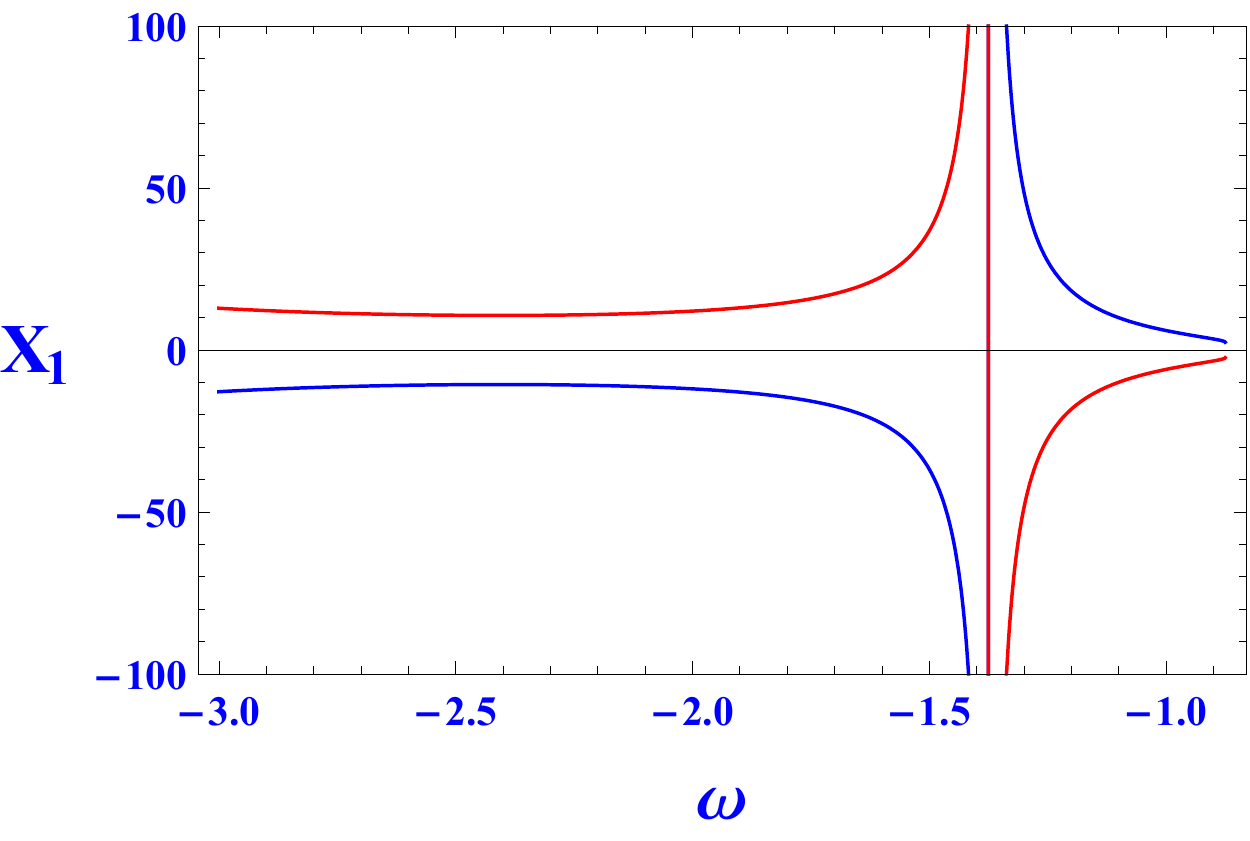}\hspace{4mm}
\includegraphics[width=3.3in]{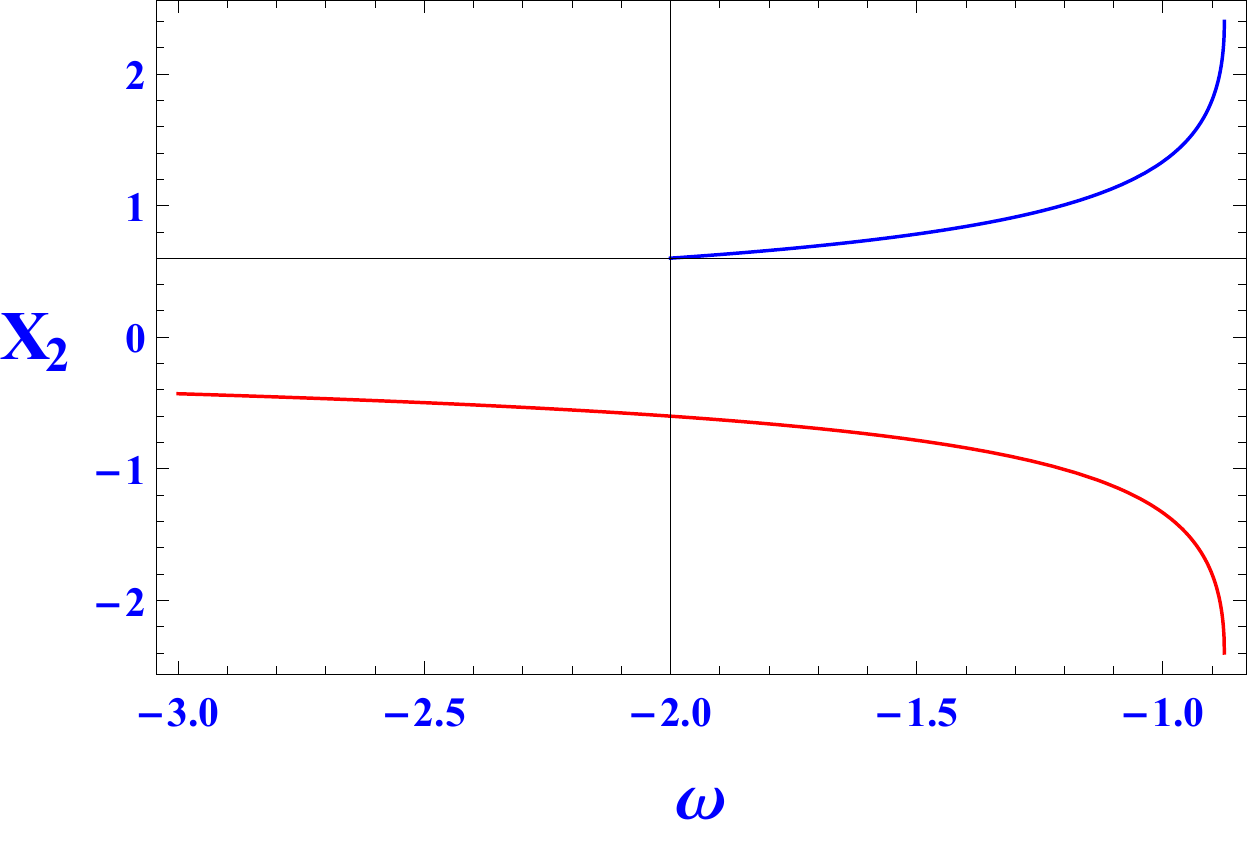}\hspace{4mm}
\caption{{\footnotesize The allowed values for $X_i$ in the $(\omega, X_i)$
parameter space associated to the $\lambda^2=1$. Positive values for $X_i$ specify
allowed values for $\omega$. The blue and red curves are
associated to the closed and open universes, respectively.
The vertical red line in the left panel corresponds to $\omega=-11/8$.}}
\foreignlanguage{english}{\label{nf-AC-alfa}}
\end{figure}

In what follows, we will investigate each case, separately, discussing the acceptable solutions associated to each case.
However, we should point out a few general features which are common to all the solutions.
The first point is that as the quantity $\chi_i$ appears in all the
solutions, so, for every acceptable solution, the BD coupling parameter must be restricted to $\omega<-7/8$.
The second point is that as $\sqrt{X_i}$ also appear in relations of $a_{i0}$,
thus, to get acceptable solutions, the right hand side of~(\ref{X}) must take positive
real values. As $X_i$ depends on the curvature
index, thus, we have to discuss about its behavior by considering also the sign of $\lambda$.\\
\\
\emph{ First solution ($i=1$):} In this case, for the closed
  universe, we find that $X_1$ takes positive values when $\omega>-11/8$,
  while for the open universe it takes positive values when $\omega<-11/8$ (see Fig.~\ref{nf-AC-alfa}, the left panel).
  Therefore, by considering the allowed range from $\chi_1$,
  we find that the acceptable range for the closed universe
  will be $-11/8<\omega<-7/8$ while for the open universe will be $\omega<-11/8$.\\
    \emph  {Second solution ($i=2$):}
  For this case, $X_2$ takes positive values when $\omega<-7/8$. Moreover, $\chi_2$ also takes real
  values for this range, thus, this range can be acceptable for the
  closed universe. However, for the open universe, $X_2$ takes negative values
  (see the right panel of Fig.~\ref{nf-AC-alfa}), thus, there is no acceptable solution for this case.\\\\
Within the above discussion, we identified allowable ranges for $\omega$.
 Now, we can apply these to find solutions associated to the open and closed universes, separately.\\

 \begin{itemize}
   \item {\bf Open Universe:} \begin{description}
                                \item[First solution:] Up to now, in order to have
                                allowable values for $\chi_1$ and $X_1$, we found
                                that the BD coupling parameter must be restricted to $\omega<-11/8$.
                                However, in this range $a^{\pm}_{10}$ and $\rho^{\pm}_{10}$ must take also positive values. The numerical results show that, in this range, $a^{+}_{10}<0$, while $a^{-}_{10}>0$, see Fig.~\ref{a01-open}, the left panel. Namely, the upper solution (plus sign) is not acceptable, while the lower solution (minus sign) can be an acceptable solution provided $\rho^{-}_{10}>0$. But just by looking at the relation of $\rho^{-}_{10}$ in (\ref{nf-dust-gen-2}), we find that $\rho^{-}_{10}$ for this range never takes only real values and it has also imaginary part. In Fig. \ref{a01-open} (the right panel), we have plotted the imaginary part of it. Thus, for this case, there is no physically acceptable solutions.
                                 \begin{figure}
\centering{}\includegraphics[width=3.3in]{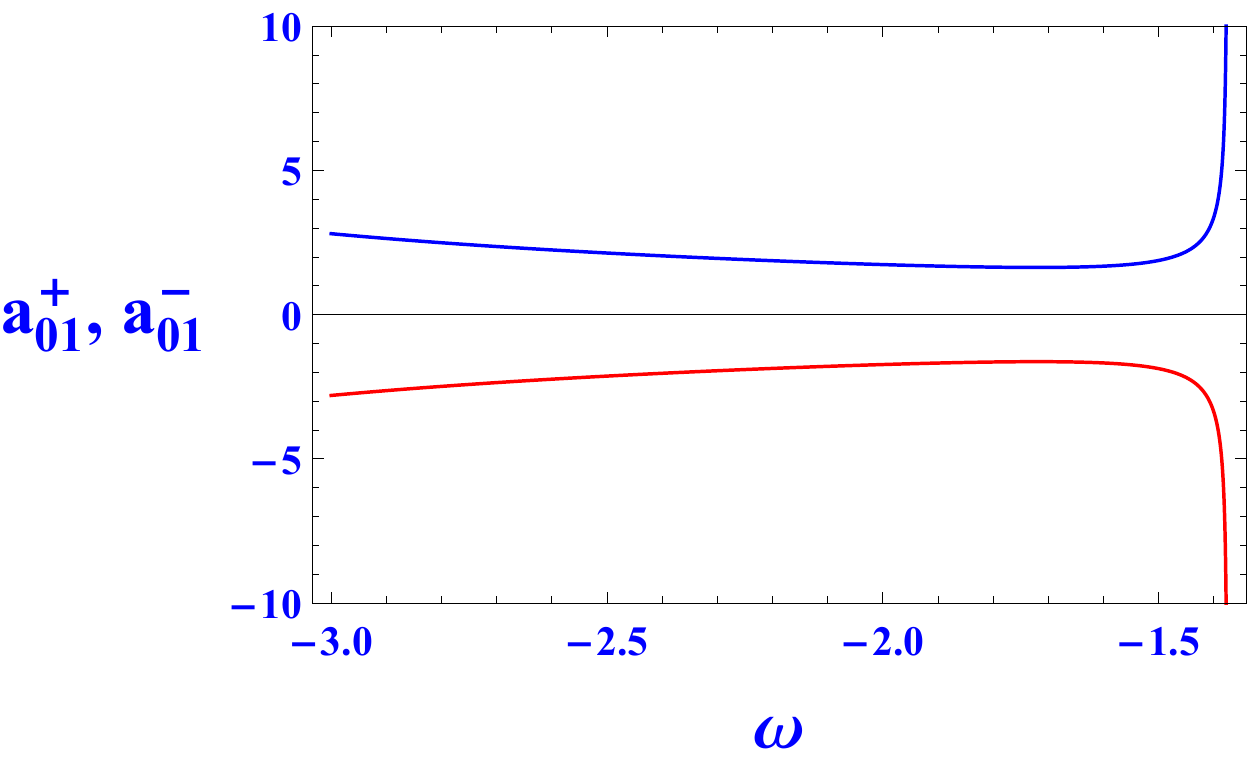}\hspace{4mm}
\centering{}\includegraphics[width=3.3in]{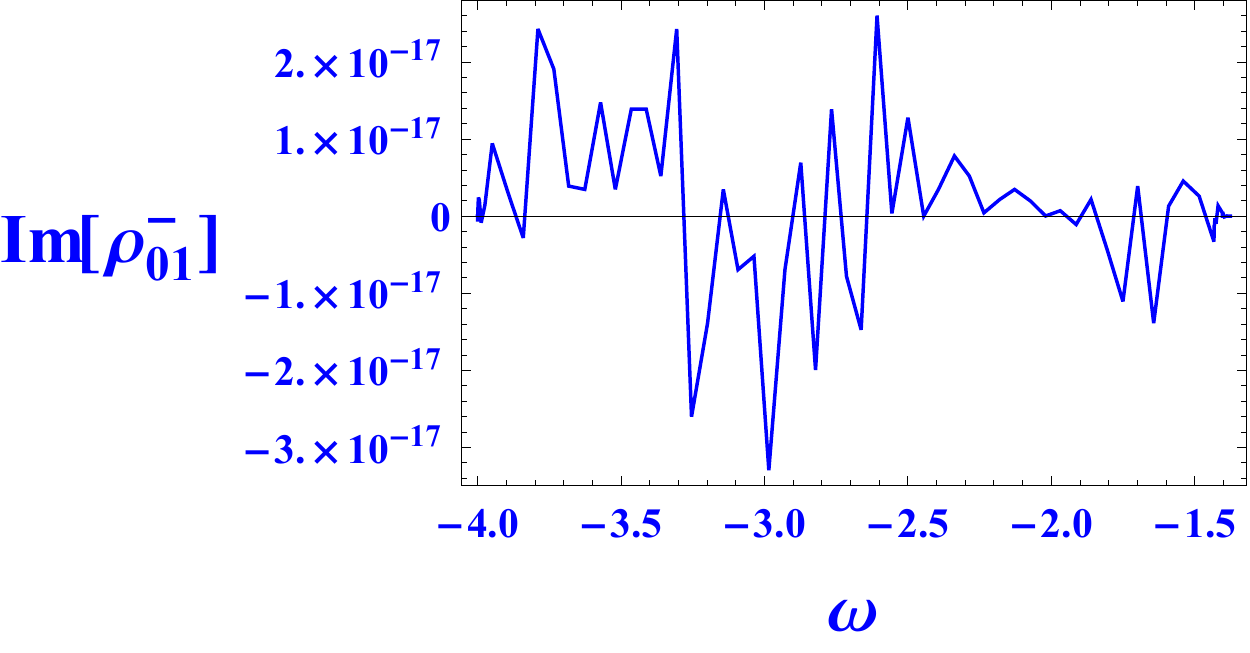}\hspace{4mm}
\caption{{\footnotesize  The left and right panels show respectively the
behavior of scale factor and the imaginary part of $\rho^{-}_{10}$ versus $\omega$.
The mentioned quantities are associated to the first
solution with $\lambda=-1$ which have been plotted in an arbitrary fixed time.
In the left panel, the blue and red curves correspond to $a^-_{01}$ and $a^+_{01}$, respectively.}}
\foreignlanguage{english}{\label{a01-open}}
\end{figure}

                              \item[Second solution:]As mentioned before, $X_2<0$, there is not any acceptable solution for this case.
                              \end{description}
 Therefore, the reduced cosmology arisen from the MBDT field equations, for the FLRW open universe which is
 filled with dust fluid just gives a few mathematical solutions which cannot be physically acceptable solutions.                             \item {\bf Closed Universe:}\begin{description}
                                 \item[First solution:] As discussed above, for this case, the allowed range is $-11/8<\omega<-7/8$.
                                 Now, we should examine this range to see whether it can generates positive real values for $\rho^{\pm}_{10}$ and $a^{\pm}_{10}$ or not. For the upper case (plus sign), we see that the range $-1<\omega<-7/8$ gives $a^{+}_{10}>0$. We can see that, for this range, $\rho^{+}_{10}>0$. In Fig.~\ref{a10-pho10} (left panel), we have plotted the behavior of $a^{+}_{10}$ and $\rho^{+}_{10}$ versus $\omega$. Our numerical results show that, in this allowable range, $\rho^{+}_{10}$ always takes very small real values which can be physically acceptable. From~(\ref{nf-dust-gen-2}), as $C>0$ and $\alpha>0$, thus, $\phi^{+}_{10}>0$ and $\psi^{+}_{10}>0$. Finally, we conclude that the first solution (upper case) in the range $-1<\omega<-7/8$ is an acceptable solution for the closed universe. \\
                                 However, for the lower case, although in the range $-11/8<\omega<-7/8$, we get $a^{-}_{10}>0$; but in this range, $\rho^{-}_{10}$ does not take real values, thus, we cannot accept this case as a physical solution.
                                 \\
                                 Let us investigate the behavior of the quantities for the upper case in terms of cosmic time. From relations
                                 (\ref{gen-solution}), the lower relation of (\ref{pot.gen-t}) and (\ref{ro-BD}), we see that $a(t)$ increases versus time, linearly;  the behavior of $\beta_1-4$, $\beta_1-2$ and $-\beta_1$ versus the allowable $\omega$ determines the behavior of the quantities associated to this case. For $-1<\omega<-7/8$, we find that $\beta_1-2$ and $\beta_1-4$ take positive real values, whilst $-\beta_1$ always takes negative values. Namely, $\phi^+_1(t)$ and $\rho_{_{\rm BD1}}^{+}$ increase versus time, while $\psi_1^{+}$ decreases versus time. As the fifth dimension contracts with the cosmic time, it is favorable. Moreover, as the BD coupling parameter increases versus time, thus the gravitational coupling decreases with time which is in agreement with the Dirac's hypothesis.
                                 \begin{figure}
\centering{}\includegraphics[width=3.3in]{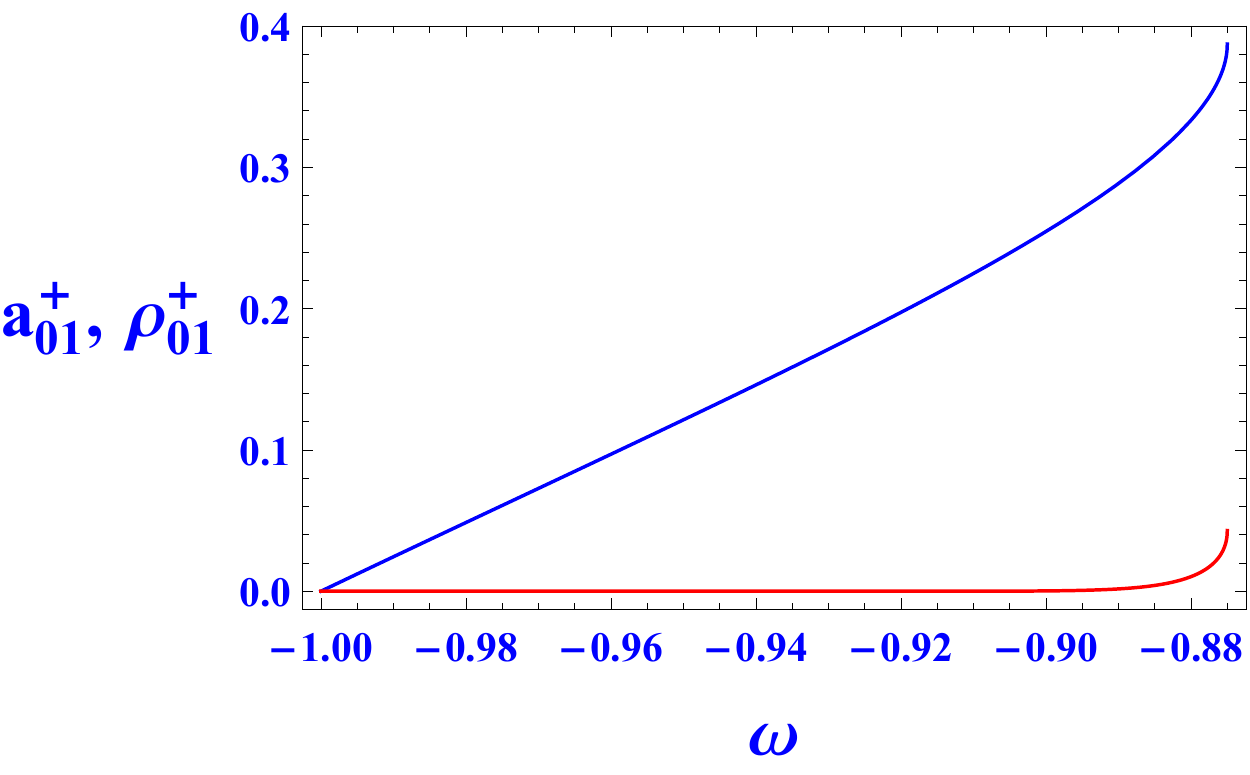}\hspace{4mm}
\centering{}\includegraphics[width=3.3in]{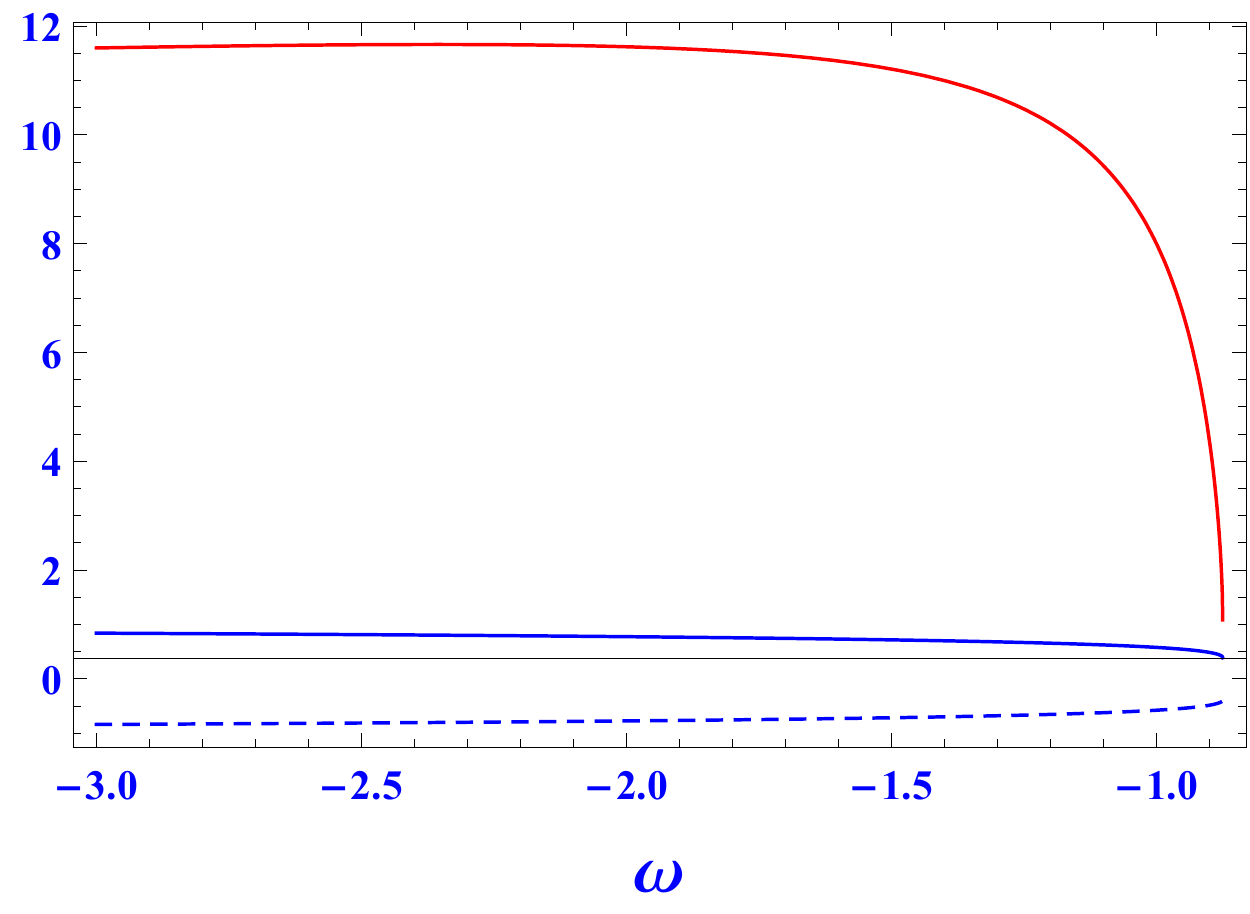}\hspace{4mm}
\caption{{\footnotesize Left panel:the behavior of the scale factor (the blue curve) and induced
energy density (the red curve) versus $\omega$ in an arbitrary fixed time
associated to the first solution (upper case) and $\lambda=1$.
Note that, regardless of the left panel, our other numerical endeavors show that
 $\rho^{+}_{10}$ at the mentioned allowed range of $\omega$
takes very small values which are nonzero.
Right panel: the behavior of $a_{02}^+$ (the blue solid curve), $a_{02}^-$ (the blue dashed curve)}
and $\rho_{02}^+$ (the red curve) versus $\omega$ associated to the second solution for the closed universe.}
\foreignlanguage{english}{\label{a10-pho10}}
\end{figure}

\item[Second solution:] For this case by considering the allowable
values for $\chi_2$ and $\sqrt{X_2}$ , we obtained the allowable
range\footnote{We should note that as when $\omega=-1$, $\beta_2$
goes to infinity, thus we exclude the value $\omega=-1$ from this solution.} $\omega<-7/8$.
Let us prob this range and investigate if positive real
values for the scale factor and the induced energy density in an arbitrary
fixed time can be obtained. From (\ref{nf-dust-gen-2}), for $\omega<-7/8$, we
find  that $a_{20}^{+}>0$ while $a_{20}^{-}<0$, see the right panel of Fig. \ref{a10-pho10}.
Namely, the lower case does not give acceptable solution. Now, we should determine
the behavior of the induced energy density for the upper case. By means of numerical
methods, it is straightforward to show that $\rho_{20}^{+}$ takes positive real
values in the allowable range, see the right panel of Fig.~\ref{a10-pho10}.

    As we have shown that the second solution (upper case) is an acceptable
    solution, determining the behavior of the quantities versus time will be a worthwhile task.
    Similar to the previous case, $a^{+}_2(t)$ increases linearly with time; and the behavior
    of $\beta_2-2$, $\beta_2-4$ and $-\beta_2$ versus $\omega$ determines the behavior of the other quantities versus time.
    We can easily show that for $\omega<-1$, $\beta_2-4$ takes negative
    values, while for $-1<\omega<-7/8$, it takes positive values. Namely,
    for the former range the induced energy density decreases by time,
    whereas for the latter range it increases with time. However, $\beta_2-4$ for all
    the values which are taken from $\omega<-7/8$ ($\omega\neq-1$) produces
    positive real values which leads to an increasing BD scalar field by the
    cosmic time. Thus, the gravitational coupling decreases with time which is again in
    agreement with the Dirac's hypothesis. Finally, as $\beta_2$ always takes positive
    values, thus the fifth dimension contracts by the cosmic time.

\end{description}
 \end{itemize}

\subsection{Radiation Cosmologies}
Radiation dominated universe in the context of the standard BD
theory has been investigated in~\cite{RF76,LP-rev,SD89,XY93,BP97}.
In what follows, we will find the specific solutions associated to radiation-dominated
universe in MBDT and discuss the properties of corresponding quantities
and their differences with those obtained from standard BD theory.
\subsubsection{\bf Flat Space}
\label{flat-rad}
By setting $W_{_{\rm BD}}=1/3$ in the upper relation of (\ref{W-BD}) and solving this
 equation together with the one obtained for the flat space in 5D, i.e., equation
 (\ref{Gen-eq-flat}), we get\footnote{The other solution is $n=0$, $\beta=1$ in which the
 BD scalar field takes constant value. To find the exact solution for this case, we must
 start from equations~(\ref{ohanlon-eq-1})-(\ref{ohanlon-eq-4}). Solving them by
 assuming a power-law for the scale factor gives $a(t)=a_0t^{1/2}$ and
 $\psi(t)=\psi_0t^{-1}$ (where $a_0$ and $\psi_0$ are constants), which is
  the unique solution of the Einstein field equations in vacuum for the spatially
 flat FLRW metric in a five-dimensional space-time.}
 \begin{eqnarray}\label{f-Beta-rad-rel}
\beta=\frac{3+2\omega}{2(1+\omega)},\hspace{10mm} n=\frac{1}{1+\omega}.
\end{eqnarray}
By substituting the above values of $n$ and $\beta$ into relations~(\ref{Dirac}), (\ref{assumption-2}),
 (\ref{gen-scaleFactor}), as well as the upper relation of (\ref{pot.gen-t}) and (\ref{ro-BD-flat}), we get the
following solutions
\begin{eqnarray}\label{flat-rad-sol}
a(t)=a_0t^{\frac{2(1+\omega)}{5+4\omega}}, \hspace{5mm}
\phi(t)=\phi_0t^{\frac{2}{5+4\omega}}, \hspace{5mm}
V&=&V_0t^{-\frac{8(1+\omega)}{5+4\omega}}, \hspace{5mm}
\rho_{_{\rm BD}}=\rho_0t^{-\frac{8(1+\omega)}{5+4\omega}},
\end{eqnarray}
where the constants $a_0$, $\phi_0$, $V_0$, and $\rho_0$ are given by
\begin{eqnarray}\label{flat-rad-sol-a}
a_0&=&\left[\frac{A(5+4\omega)}{2C\alpha}\right]^{\frac{2(1+\omega)}{5+4\omega}},\\
\label{flat-rad-sol-phi}
\phi_0&=&C\left[\frac{A(5+4\omega)}{2C\alpha}\right]^{\frac{2}{5+4\omega}},\\
\label{flat-rad-sol-V}
V_0&=&\frac{C(3+2\omega)}{4}\left(\frac{A}{2C\alpha}\right)
^{\frac{2}{5+4\omega}}(5+4\omega)^{-\frac{8(1+\omega)}{5+4\omega}},\\
\label{flat-rad-sol-rho}
\rho_0&=&\frac{3A(3+2\omega)}{32\pi\alpha}
\Big[\frac{A(5+4\omega)}{2C\alpha}\Big]^{-\frac{(3+4\omega)}{5+4\omega}},
\end{eqnarray}
where the constants $A$, $C$ and $\alpha$ can be determined by knowing the age of the
universe and its energy density at the present time.
The scale factor of the fifth dimension is given by
\begin{eqnarray}\label{psi-rad-rel}
\psi(t)&=&\alpha\left[\frac{A(5+4\omega)t}{2C\alpha}\right]^{-\frac{3+2\omega}{5+4\omega}}.
\end{eqnarray}
Let us first find the allowable ranges of $\omega$ for
this case. As $a_0$, $\phi_0$, $\psi_0$ and $\rho_0$ should take positive real values,
from relations~(\ref{flat-rad-sol-a}) and~(\ref{flat-rad-sol-rho}), we find
that the allowed ranges of the BD coupling parameter, which depends on the sign of $A$.
Therefore, in what follows, we would investigate two separated solutions.
\begin{description}
  \item[{\bf Case I; $A<0$:}] In this case, when the BD coupling parameter is restricted to $\omega<-5/4$, the
  quantities $a_0$, $\phi_0$ and $\psi_0$ take positive real values
   whereas $\rho_0$ takes positive real values when $\omega<-3/2$. Therefore,
   the allowed range for this case is $\omega<-3/2$; see the upper panels of Fig.~\ref{q-flat-rad}.
   For this allowed range, it is straightforward to
   show that $0<2(1+\omega)/(5+4\omega)<1$.
   More precisely, the scale factor of the universe decelerates with cosmic time.
   Furthermore, the induced energy density, BD scalar field and the fifth dimension decrease with time.
     \item[{\bf Case II; $A>0$:}]
In this case, to get positive real values for the
corresponding quantities (in an arbitrary fixed time), $\omega$
 must be restricted to $\omega>-5/4$ (see the lower panels of Fig.~\ref{q-flat-rad}),
 which gives reasonable results for the radiation-dominated universe as follows.
 It is easy to show that
(i) for $-5/4<\omega<-1$, we get $2(1+\omega)/(5+4\omega)<0$, while for $\omega>-1$, we get $0<2(1+\omega)/(5+4\omega)<1$.
The former range (of the power of the cosmic time associated to the scale factor)
leads us to a spatially flat universe whose scale factor contracts
with the cosmic time; whereas the latter leads to a decelerating universe.
(ii) the energy density increases with the cosmic time for $-5/4<\omega<-1$
whereas it decreases with time when $\omega>-1$.
(iii) the BD scalar field increases with the cosmic time; in this case the gravitational
coupling decreases with time which is in agreement with Dirac's hypothesis.
(iv) $\psi(t)$ also decreases with cosmic time for all allowed values of $\omega$.
The above properties imply that the MBDT scenario, by assuming
the mentioned constraint on the BD coupling parameter, can be a
relevant model for getting a radiation-dominated universe when $\omega>-1$, in which
the induced matter can play properly the role of ordinary matter in the universe.

 \begin{figure}
\includegraphics[width=3.3in]{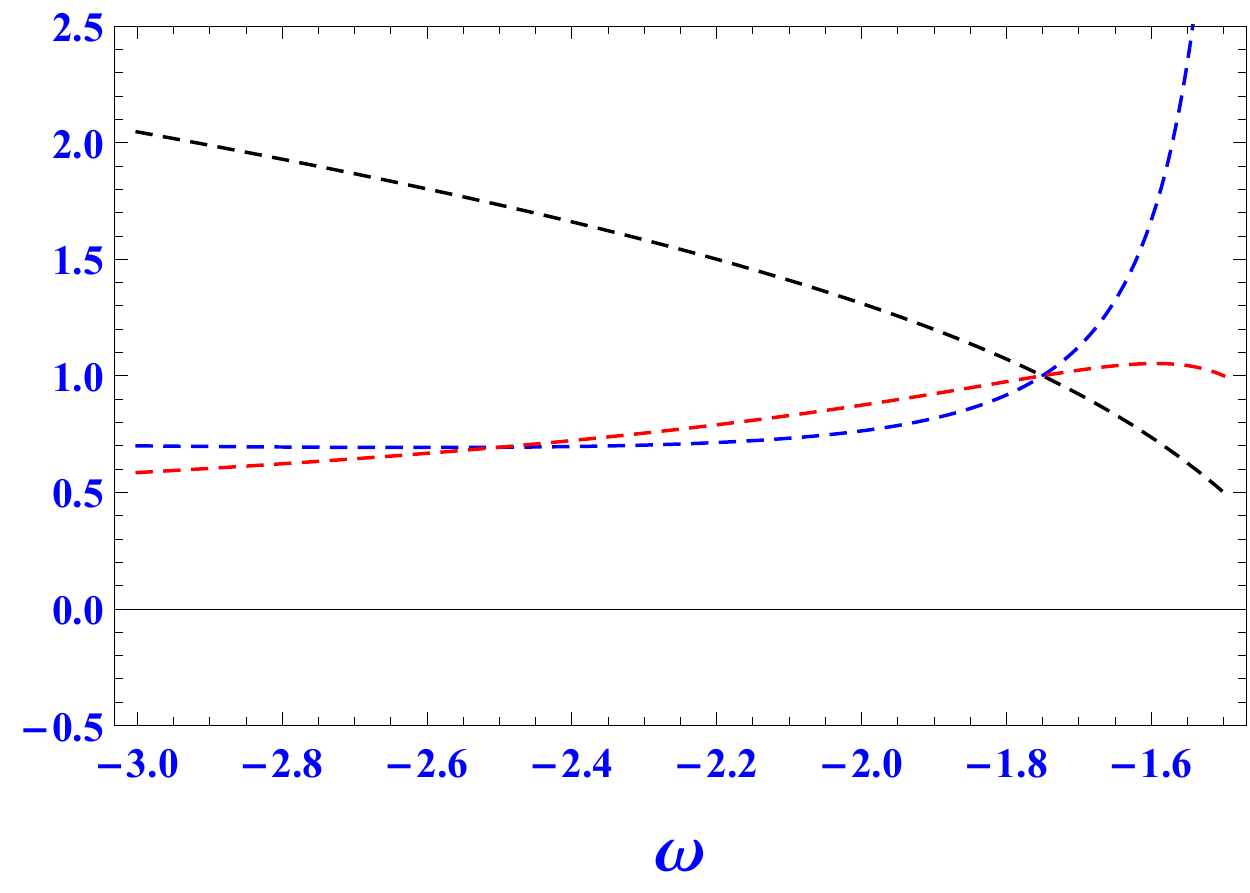}\hspace{4mm}
\includegraphics[width=3.3in]{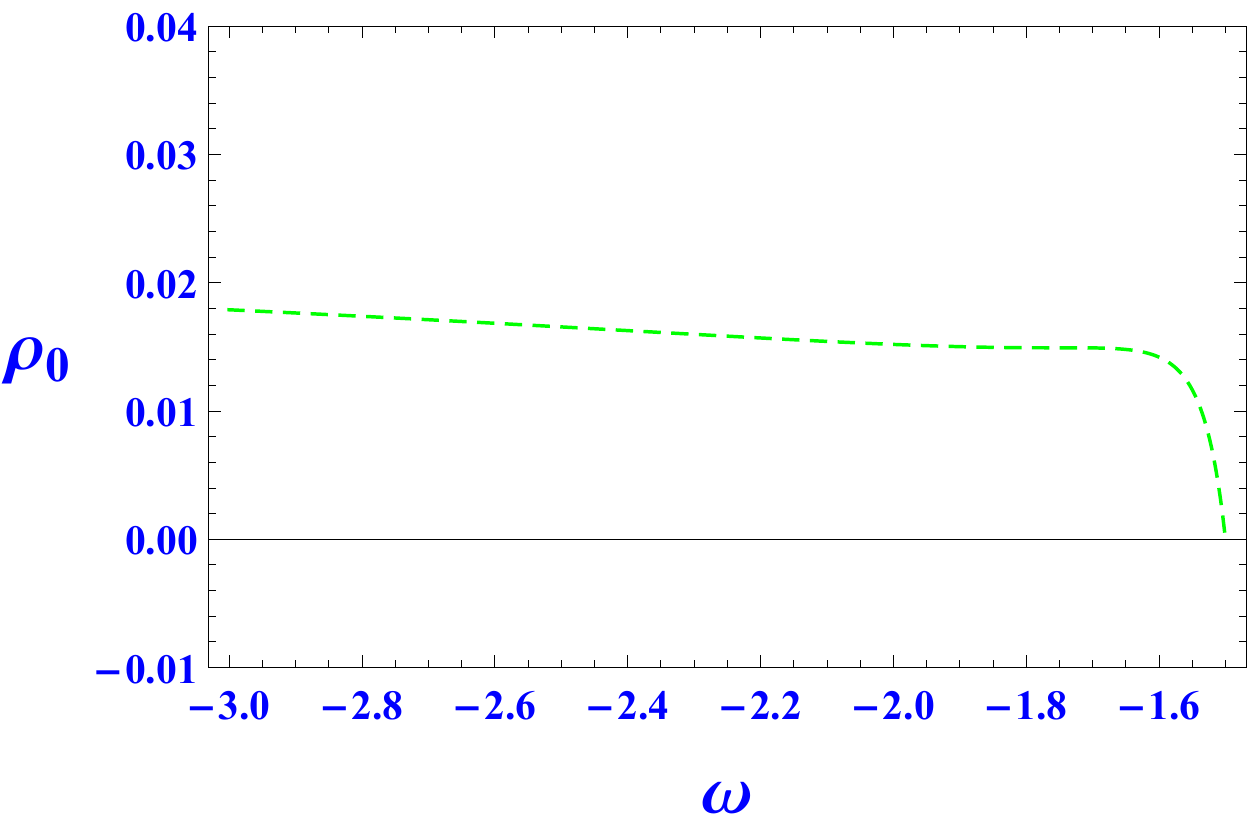}\hspace{4mm}
\includegraphics[width=3.3in]{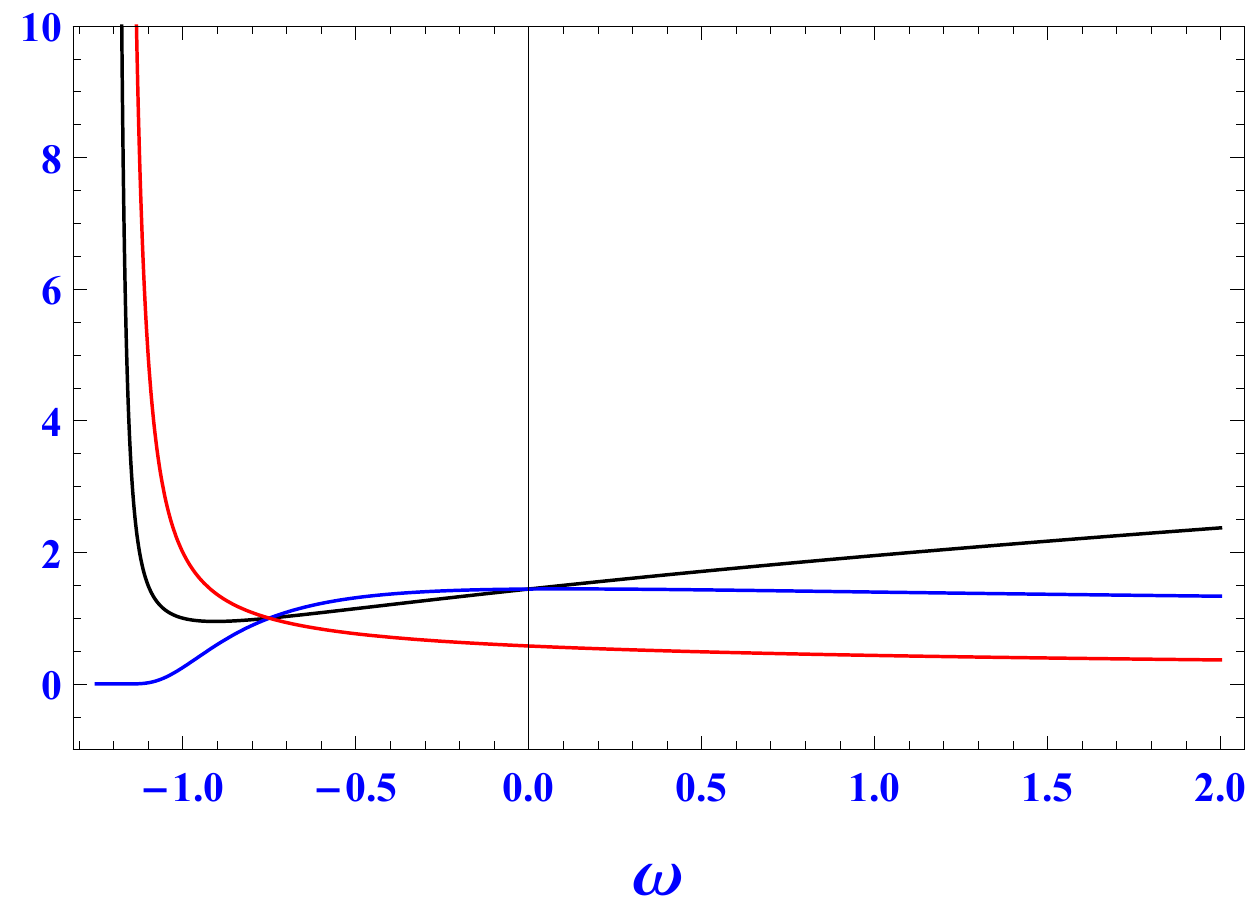}\hspace{4mm}
\includegraphics[width=3.3in]{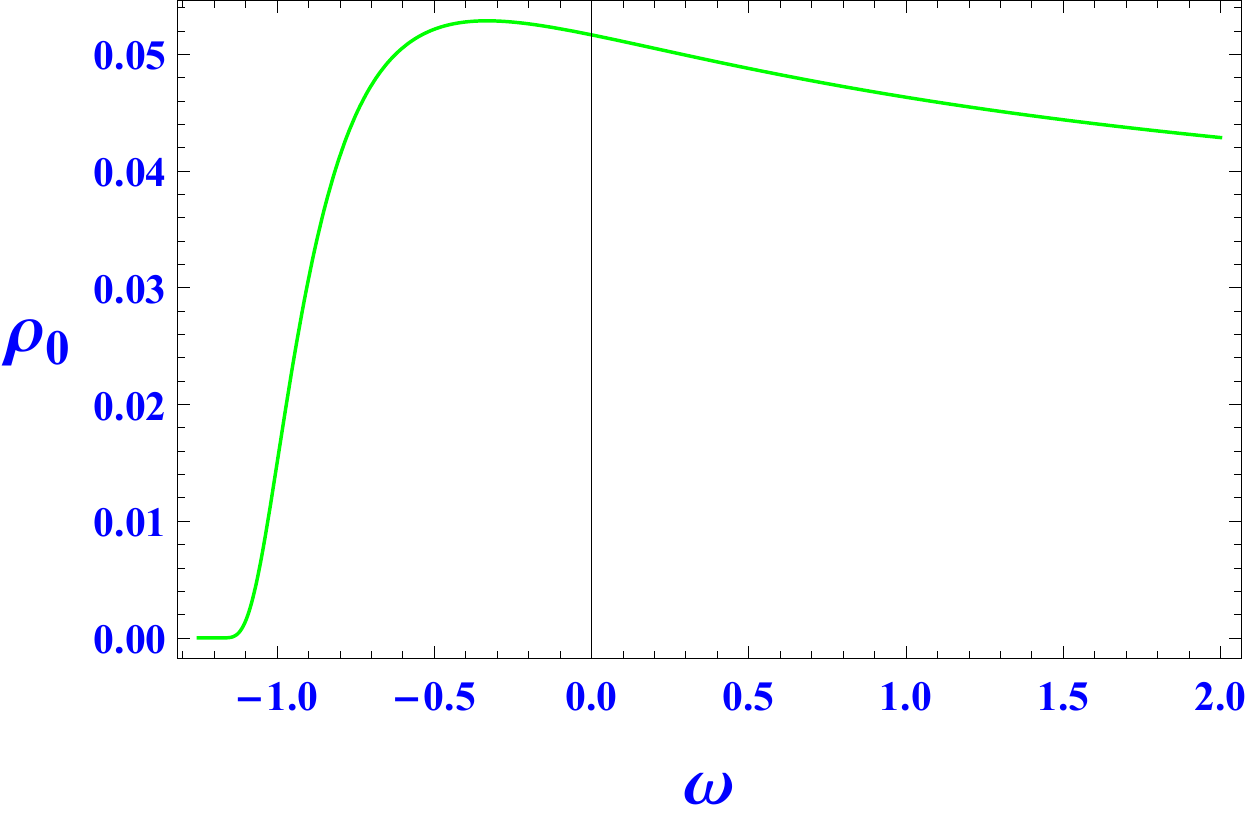}\hspace{4mm}
\caption{{\footnotesize The behavior of $a$ (black curves), $\phi$ (the blue curves), $\psi$ (the red curves)
and $\rho_{_{\rm BD}}$ (the green curves) versus $\omega$ (in an arbitrary fixed time) for the spatially
flat FLRW universe which is filled with radiation.
The dashed and solid curves are associated to $A<0$ and $A>0$, respectively.
We have chosen the constants as $A=\pm1$ and
$C=\alpha=1$ in agreement with the assumptions of cases I and II.}}
\foreignlanguage{english}{\label{q-flat-rad}}
\end{figure}

\end{description}

\subsubsection{{\bf non-flat space}}

For this case, by setting $W_{_{\rm BD}}=1/3$ in the lower
relation of~(\ref{W-BD}), it is easy to show that\footnote{The other solution for
this case is $\beta=2$, which, according to (\ref{gen-solution}), yields a constant value
for the BD scalar field and infinite value for the scale factor.
As discussed in section~\ref{5d-cosmology}, in this case the BD theory may
reduce to GR and thus such a result is reasonable.}
\begin{eqnarray}\label{Beta-rad-rel}
\beta=\frac{4(3+2\omega)}{5+4\omega},
\end{eqnarray}
where $\omega\neq-5/4$. \\

By substituting $\beta$ from (\ref{Beta-rad-rel}) into (\ref{Gen-AC-alpha}), we get
\begin{equation}\label{Rad-AC-alpha}
\left(\frac{A}{C\alpha}\right)^2=\frac{4\lambda}{(3+2\omega)(11+8\omega)},
\end{equation}
where $\omega\neq-3/2,-11/8$.
By employing relations (\ref{Beta-rad-rel}) and (\ref{Rad-AC-alpha}) for the general solutions
associated to the non-flat space, namely (\ref{gen-solution}), we get
\begin{equation}\label{Rad-nf-gen}
a^{\pm}=a_0^{\pm}t,\hspace{5mm}\phi^{\pm}=\phi_0^{\pm}t^{\frac{2}{5+4\omega}},
\hspace{5mm}\psi^{\pm}=\psi_0^{\pm}t^{-\frac{4(3+2\omega)}{5+4\omega}},
\hspace{5mm}\rho_{_{\rm BD}}^{\pm}=\rho_0^{\pm}t^{-\frac{8(1+\omega)}{5+4\omega}},V^{\pm}=V_0^{\pm}t^{-\frac{8(1+\omega)}{5+4\omega}}
\end{equation}
where
\begin{eqnarray}\label{Rad-nf-rel-1}
a_0^{\pm}&=&\pm(5+4\omega)\sqrt{\frac{\lambda}{(3+2\omega)(11+8\omega)}},\\
\label{Rad-nf-rel-2}
\phi_0^{\pm}&=&C\left(a_0^{\pm}\right)^{\frac{2}{5+4\omega}},\hspace{5mm}\psi_0^{\pm}=\alpha\left(a_0^{\pm}\right)^{-\frac{4(3+2\omega)}{5+4\omega}}\\
\label{Rad-nf-rel-3}
\rho_0^{\pm}&=&\frac{3C}{16\pi}(3+2\omega)(11+8\omega)\left(\frac{2}{5+4\omega}\right)^{\frac{8(1+\omega)}{5+4\omega}}\left[\pm\sqrt{\frac{4\lambda}
{(3+2\omega)(11+8\omega)}}\right]^{\frac{2}{5+4\omega}},\\\\
\label{Rad-nf-rel-4}
V_0^{\pm}&=&C(3+2\omega)\left(\frac{2}{5+4\omega}\right)^{\frac{8(1+\omega)}{5+4\omega}}\left[\pm\sqrt{\frac{4\lambda}
{(3+2\omega)(11+8\omega)}}\right]^{\frac{2}{5+4\omega}},
\end{eqnarray}

In what follows, we discuss
the behavior of the quantities for closed and open Universes, separately.\\
\begin{itemize}
  \item
{\bf Open Universe:} For this case, by substituting $\lambda=-1$ into~(\ref{Rad-nf-rel-1}), we find that, in order to
get real values for the square root, the BD coupling parameter must be restricted to $-3/2<\omega<-11/8$.
On the other hand, $a_0$ must take positive values, thus for the upper case (plus sign)
we get $\omega>-5/4$, which does not have overlap with the previous allowable range of $\omega$.
Therefore, there is no consistent physical solution for the upper case.
However, for the lower case (minus sign), $\omega$ must be restricted to $\omega<-5/4$, which has a overlap with the previous allowed range.
Namely, in this case, the range $-3/2<\omega<-11/8$ not only gives a
real values for $A/C\alpha$ but also produces positive values for $a_0$.
Nonetheless, this is not enough to have a acceptable physical solution. More concretely, other
quantities such as $\rho_0^{-}$, $\phi_0^{-}$ and $\psi_0^{-}$ must take real positive values
in the mentioned range of $\omega$.
However, just by investigating $\rho_0^{-}$, we find that it takes imaginary values in the mentioned range.
Finally, we can conclude that the set of solutions (\ref{Rad-nf-gen}) for an open universe are only
valid as mathematical solutions, namely, for the FLRW universe
with $\lambda=-1$ (when $\phi$ and $\psi$ are related to the scale factor
with a power-law equation), the MBDT scenario does not yield any physical solution.
\item
{\bf Closed Universe:} By substituting $\lambda=1$ into relations
(\ref{Rad-nf-gen})-(\ref{Rad-nf-rel-4}) we get two classes of solutions for this case.
Let us describe the corresponding quantities for this case by finding the allowable range for the BD coupling parameter.
From (\ref{Rad-nf-rel-1}), we see that to have real values for the square
root, $\omega$ must be restricted to either $\omega<-3/2$ or $\omega>-11/8$. By considering these
allowable ranges of $\omega$, in what follows, we investigate
the solutions associated to the upper case (plus sign) and lower case (minus sign), separately.
\begin{description}
  \item[{\bf Upper sign:}]
   In this case, from (\ref{Rad-nf-rel-1}), we find that $a_0^+$ takes positive values if the BD
  coupling parameter is restricted to $\omega>-5/4$. For this allowed range of $\omega$,
  as $C>0$ and $\alpha>$, thus, $\rho_{_{\rm BD}}$, $\phi$ and $\psi$ take positive real
  values for an arbitrary fixed time, see Fig. \ref{nf-rad-clos-up}. Consequently, the acceptable
  physical solutions associated for a closed universe with upper sign (which is filled with radiation)
  are given by (\ref{Rad-nf-gen}) in which $\omega$ must be larger than $\omega>-5/4$.
  \begin{figure}
  \centering{}\includegraphics[width=3.3in]{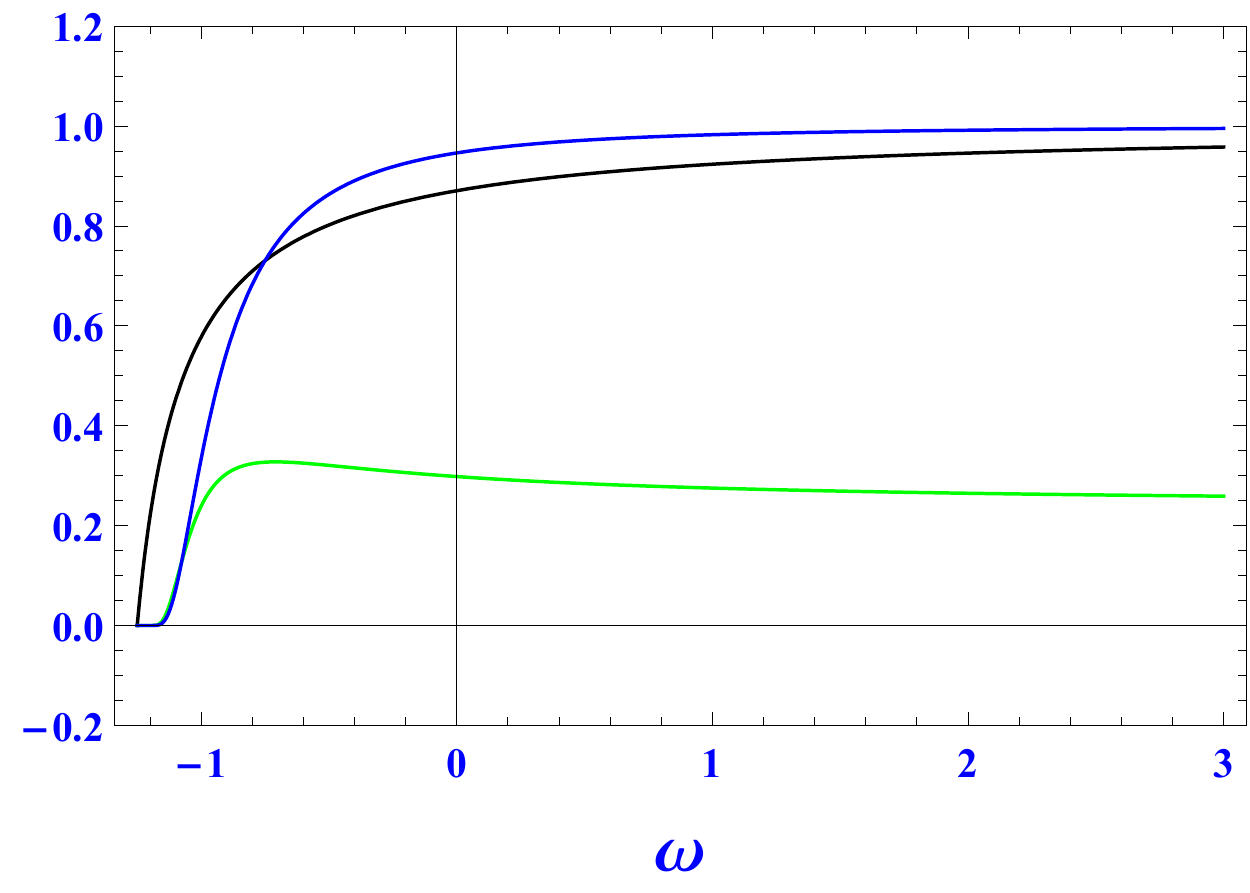}\hspace{4mm}
\centering{}\includegraphics[width=3.3in]{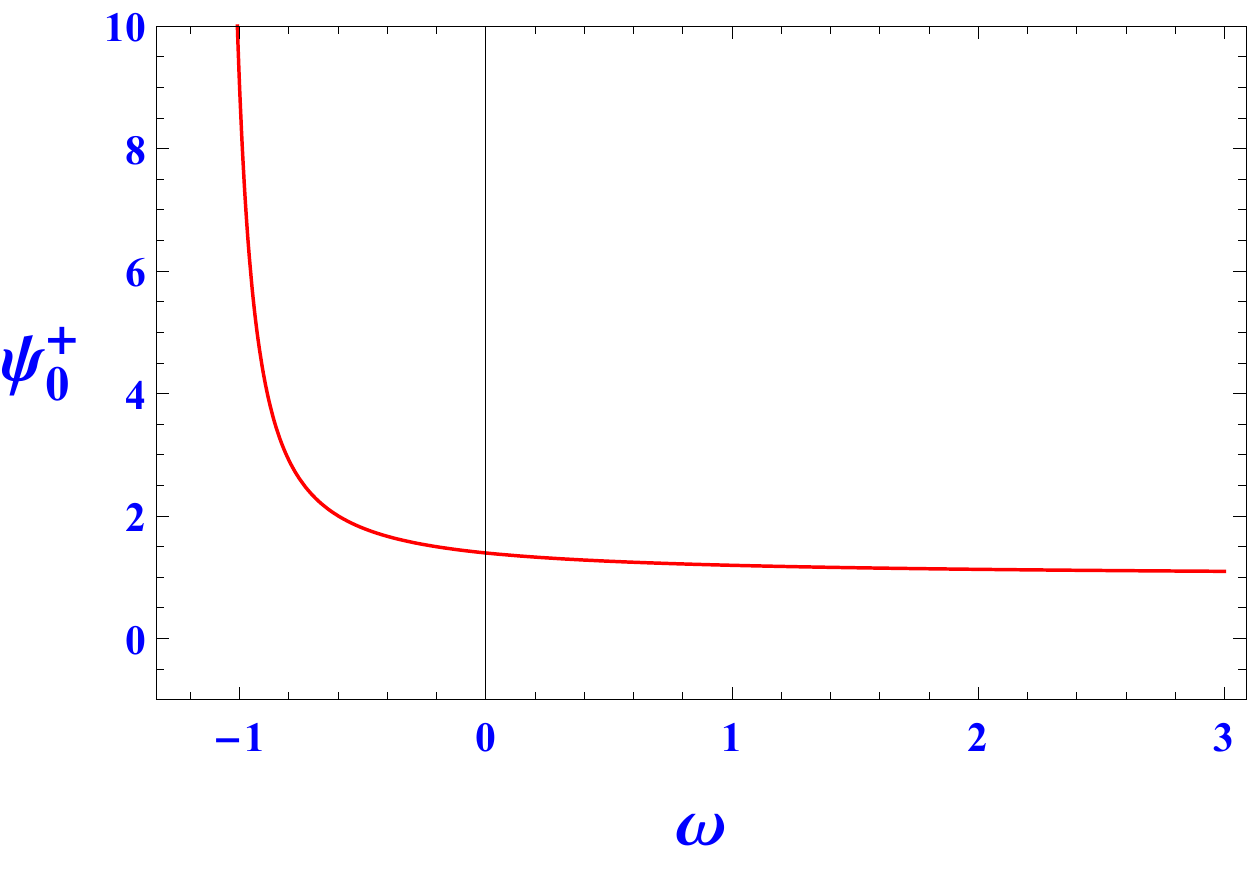}\hspace{4mm}
\caption{{\footnotesize The behavior of $a$ (the black curve), $\rho_{_{\rm BD}}$ (the green curve), $\phi$ (the blue curve)
and $\psi$ (the red curve in the left panel) in an arbitrary fixed time versus $\omega$ ($\omega>-5/4$)
associated to the closed universe (and upper sign) which is filled with radiation.
We have chosen $C=1=\alpha$. Note that $\phi_0^+$ and $\rho_0^+$ take
 very small (but nonzero) values when $\omega\rightarrow-5/4$.}}
\foreignlanguage{english}{\label{nf-rad-clos-up}}
\end{figure}
It is also noteworthy to describe the time behavior of quantities associated to this case.
From (\ref{Rad-nf-gen}), it is straightforward to show that for $\omega>-5/4$,
(i) as $\phi_0^+>0$ and $2/(5+4\omega)>0$, thus, the BD scalar field always
increases with the cosmic time; hence, the gravitational
coupling decreases by cosmic time which is in agreement with Dirac's hypothesis;
(ii) as $\rho_0^+>0$, we see that the induced energy density increases with
time when $-5/4<\omega<-1$, whilst it decreases with time when $\omega>-1$;
(iii) $\psi^+(t)$ always decreases with the cosmic time which is a relevant outcome.

\item[{\bf Lower case:}] For this case, $a_0^-$ takes positive values when the BD coupling
  parameter is restricted to either $-11/8<\omega<-5/4$ or $\omega<-3/2$.
  However, likewise for the lower case for the open
  universe, we cannot obtain real values for the energy density in arbitrary fixed time when $-11/8<\omega<-5/4$.

  For the other range, i.e., $\omega<-3/2$, from (\ref{Rad-nf-rel-2}), we
  see that (by assuming $C>0$ and $\alpha>0$), $\psi^-$ and $\phi^-$ always take positive real values.
  Whereas, from (\ref{Rad-nf-rel-4}), we find that $\rho_0^-$ does not take real values.
  Thus, this case is not a physically acceptable solution.
\end{description}
\end{itemize}


\section{summary and discussions}
\label{concl.}
The MBDT~\cite{RFM14} has four sets of field equations: one set corresponds to
the generalized version of the conservation law introduced in IMT~\cite{PW92,stm99}.
Another set is a nonlinear differential equation associated to
the scale factor of the extra dimension which has no analog in the standard BD theory.
Finally, the other two sets can be related to the conventional BD action, but with specific scalar potential, in
four dimensions in which the matter and the scalar potential have an intrinsic geometrical origin.
This induced matter is composed of three parts, see Eq. (\ref{matt.def}).
The first part is a function of the first and second derivatives of
the metric components with respect to the extra dimension.
The second and third parts are functions of the BD scalar field, metric
components and their derivatives with respect to the extra dimension.

Achieving unification of matter and geometry has
been claimed as the main motivation for introducing a large extra dimension in IMT~\cite{Pon01}.
As the MBDT can be considered as an extended version of the IMT, thus
 the motivation that has been followed to also consider a large extra dimension to construct the MBDT scenario.
 However, in MBDT, in addition to the geometrically induced matter,
 there is also an induced scalar potential
 which has also been of interest to discuss in MBDT.
 This scalar potential has been employed to yield either an
 accelerating universe~\cite{Ponce1,Ponce2,RFM14} (by assuming a
 spatially flat FLRW universe) or obtaining more general solutions
 for a Bianchi type I model~\cite{RFS11}.
 The main objective of our herein work was to employ the MBDT scenario to obtain new extended
 solutions associated to both the spatially flat FLRW
 universe (by assuming more generalized power-law solutions than ones assumed in the previous works) and
 the non-flat space with respect the corresponding
 solutions in the context of the conventional BD theory.

  In this paper,
 we started from the geometry of the five-dimensional bulk
 and then constructed the physics on the projected four-dimensional hypersurface. More precisely,
 by considering a five dimensional FLRW universe
 (without ordinary matter) with all values of the curvature index,
 we have derived the equations of the standard BD theory whose
 scalar field only depends on the cosmic time. Then, by assuming the Dirac's
 hypothesis, which claims that the gravitational constant
 should be connected to the scale factor of the universe in a power-law relation,
 we have solved the equations
 of motion. Our results show that there is a general solution for the non-flat
 space [see Eqs.~(\ref{gen-solution})] and two kinds of solutions for the flat
 space, the general power-law solution and exponential solution.
 We have also discussed regarding some particular cases of theses
 solutions when either $\omega$ or $\beta$ takes special values.

Subsequently, we have employed the MBDT set up (reviewed in section~\ref{Set up})
to construct the physics on a four-dimensional space-time.
First, we found the relations associated to the induced matter and induced
scalar potential in general cases. Our results have shown that the induced scalar potential
is in either logarithmic or power-law forms.
As the former leads to an inconsistency to apply the energy
conditions, we abstained from proceeding to study it.
However, the latter yielded the equations of state for the barotropic matter for all
values of the curvature index. Such an induced matter,
in which all the terms emerge from the geometry of the fifth dimension, has interesting
properties. Namely, it obeys the conservation law similar to the ordinary matter in the standard
BD theory. Thus, it respects the Principle of Equivalence.
Besides, it is worthwhile to emphasize that the induced scalar
potential, which contributes to construct the induced matter and consequently the behavior of the BD scalar field,
 has been induced from the geometry (as a fundamental concept) rather than adding it by hand to the action.
As there is a nonzero scalar
 potential, thus, the relations associated to the scale
 factor, the BD scalar field and the components of the
 induced matter, which depend effectively on $\beta$
 and/or $\omega$, are more generalized than those of the conventional BD theory.

We proceeded to study the properties of the induced matter on
the hypersurface as consequence of the effects of the geometry of the fifth dimension and contrast
it with the ones reported from ordinary matter in the BD theory.
For that, we have concentrated on known types of the matter, which, in what
follows, we summarize and compare them with those obtained from the conventional BD theory, IMT and GR.
\begin{enumerate}
  \item {\it Vacuum solutions:} For both the flat and non-flat spaces, the only manner to get vacuum
  solutions on a four dimensional hypersurface is setting $\beta=0$, i.e. assuming $\psi={\rm constant}$.
  In this case, we found that the induced scalar potential and the components of the induced matter vanish.
  We have shown that the solutions associated to the herein model are similar
  to those obtained in the standard BD theory.
  Namely, we obtained the same exact solutions obtained by O'Hanlon-Tupper~\cite{o'hanlon-tupper-72}
  and Dehnen-Obreg\'{o}n~\cite{DO72b} in the context of the standard BD theory in vacuum.
  By means of numerical analysis, we further investigated the time behavior
  of the scale factor, BD scalar field and the gravitational constant for
  each solution. We should note that the calculations associated to the
  vacuum case can be an appropriate procedure to test the correctness
  of the results produced by the MBDT scenario.
  For all the solutions associated to the vacuum case, we see that the power-law
  assumption between the scale factor of the fifth dimension and the BD scalar
   field as well as the scale factor of the universe, is in contradiction with the Dirac's hypothesis.
    The other general consequences of these
    solutions have shown that, similarly to the conventional BD theory,
    the Birkhoff theorem and the Mach's principe, which
    implies that the matter in the universe determines the value of the gravitational
    coupling, are not valid in the context of the MBDT.
  \item {\it Dust solutions:}
  For a flat space, we found two classes of mathematical solutions
  [the upper and lower cases, see Eqs.~(\ref{flat-dust-sol-a})-(\ref{flat-dust-sol-rho})]
  which depend on the sign of integration constant $A$.
  We have discussed the effects of the sign of $A$ on the solutions associated to the
  upper and lower cases. By assuming $A<0$, we found that only one of the
  solutions (the upper case) is physically acceptable when $\omega<-4/3$.
  For this case, we have shown that the scale factor of the universe
  accelerates, while the BD scalar field and induced energy density decrease with cosmic time.
  Let us compare the result with the observational data:
  from relations~(\ref{m-dust}), it is straightforward to show that,
  when $-2<\omega<-3/2$ then $r^+$ can be restricted as $1.56\lesssim r^{+}\lesssim2.94$ or, equivalently, the deceleration
  parameter (at present time) $q_0^{+}=(1-r^+)/r^+$ is
  restricted to $-0.66\lesssim q_0^{+}\lesssim-0.36$, which is in agreement
  with the recent observational measurements~\cite{Giostri12}.

  For $A>0$, we found
  that both of the solutions can be physically acceptable.
  We have also discussed the properties of the quantities for this case.
  We found that, for both of the upper and lower solutions,
   the gravitational constant decreases with cosmic time, which is in agreement
  with Dirac's hypothesis. It is worthy that for all of the
  mentioned dust solutions associated to the flat space, the fifth dimension
  decreases with cosmic time which can be an interesting result in the context of higher dimensional models.

  Let us summarize and further discuss solutions~(\ref{f-dust-n-beta})-(\ref{psi-dust}) for large values of $\omega$.
For these cases (upper case with $A<0$ and lower case with $A>0$), without loss of generality, we can set $A=Cn\alpha$.
As $C$ and $\alpha$ always take positive values and $n$ takes negative and
positive values for the former and latter cases, respectively, thus, $A$ takes
negative and positive values for the former and latter, respectively.
From this, it is easy
to show that when $\omega\rightarrow-\infty$,
we get $a(t)=(3t/2)^{3/2}$, $\phi=C$, $\psi=(2\alpha/3)t^{-1}$ and $\rho=(C/3\pi)t^{-2}$.
Namely, the results are in agreement with those obtained from IMT and thus with GR.
  \\

  For non-flat space, $\lambda^2=1$, we got four different classes of
  mathematical solutions, see the set of relations~(\ref{nf-Dust-beta})-(\ref{nf-dust-gen-2}).
  By further probing of the properties of the solutions associated
  to the open universe, we discovered that there is no
   physically acceptable solution for it. This result is in accordance with
   the one obtained for the standard BD theory~\cite{DO71}.

   However, for the closed universe, among four classes
   of mathematical solutions, we have shown that only two of
   them can be physically acceptable. In one class of those solutions (the first solution, upper case), the BD
   coupling parameter must be restricted to $-1<\omega<-7/8$, whereas for the other
   case, we found that the allowed range is $\omega<-7/8$ (the second solution, upper case). In comparison with the corresponding
   solution in the standard BD theory (with vanishing scalar potential)~\cite{DO71},
   we find that there is no analog for the former case.
   However, in the latter case, the allowed range for $\omega$ in
   our model is replaced with $\omega<-2$~\cite{DO71}, in which probing the
   solutions associated to the large values of $\omega$ needs careful calculations~\cite{M82,M01}.
   We can further probe the results of our second solution (upper case) when $\omega\rightarrow-\infty$.
   In this case, we get $\phi=C={\rm constant}$, $a(t)=\sqrt{3/2}t$,
   $\rho(t)=C/(\pi t^{2})$ and $\psi(t)=2\alpha/(3t^2)$ which may coincide
   with the results obtained for the corresponding solutions in IMT~\cite{stm99}.
   For both the physically acceptable solutions, we have found that for the corresponding allowable ranges of $\omega$,
   the fifth dimension and the gravitational coupling decrease with the cosmic time
   and thus these solutions can be of interest.

  \item {\it Radiation solutions:} 

    For the spatially flat FLRW universe, depending on the sign of $A$,
    we got two different classes of acceptable solutions.
  The first class of the solutions corresponds to $A<0$, in which the allowed
  range is $\omega<-3/2$. For this case, the scale
  factor of the universe decelerates with cosmic time, and the BD scalar
  field, the induced energy density and the fifth dimension decrease with time.
  The second class, which corresponds to $A>0$, implies a different range of
  the BD coupling parameter, i.e., $\omega>-5/4$.
  In this case, we found that although the mentioned range gives
  physically acceptable solutions, but in the range $\omega>-1$
  we got a decelerating universe;
  the induced energy density and the fifth dimension decrease with cosmic time;
  and the BD coupling parameter increases with cosmic time which is in agreement with the Dirac's hypothesis.
  When $|\omega|$ takes large values, without loss of generality, by assuming $A=\pm Cn\alpha$
  (where $C>0$ and $\alpha>0$) in relations~(\ref{f-Beta-rad-rel})-(\ref{psi-rad-rel}), when $\omega\rightarrow\pm\infty$,
  it is straightforward to show that for both of the solutions we
  get $\phi=C={\rm constant}$, $\psi(t)=(\sqrt{2}\alpha/2)t^{-1/2}$,
   $a(t)=(2t)^{1/2}$ and $\rho=(3C/32\pi)t^{-2}$, which are in agreement
   with the solutions associated to the FLRW universe (which is filled with radiation) in IMT and thus in GR.

  For the non-flat space, we obtained two classes of mathematical solutions for each the open and closed universes.
  For an open universe, both of classes
  give neither positive nor real
  values for the induced energy density, thus, they cannot be physically acceptable solutions.
  For a closed universe, only one of the solutions (the upper case) is acceptable when $\omega>-1$.
  For this class of the solutions, we have presented the properties of the corresponding quantities.
  For instance, we have shown that the gravitational constant, the induced energy density and $\psi(t)$ decrease
  with the cosmic time, whereas the BD scalar field increase with time
  which is in agreement with the Dirac's hypothesis. All of the properties
  for this class of solutions show that it is a physically allowed cosmological model for a radiation dominated universe.
  When $\omega\rightarrow\infty$, from relations (\ref{Beta-rad-rel})-(\ref{Rad-nf-rel-4}), we obtain
  $\phi=C={\rm constant}$, $a(t)=t$, $\psi(t)=\alpha t^{-2}$ and $\rho(t)=(3C/4\pi)t^{-2}$,
  which are in agreement with the corresponding ones obtained in IMT.
\end{enumerate}
Finally, it is worthwhile to mention a few points regarding the MBDT setting~\cite{RFM14}
 and the obtained solutions of our herein work:
\begin{itemize}
  \item  Perhaps it would be a good idea to presume the MBDT as a
  powerful and a fundamental setting to obtain the
exact solutions which correspond to those of the standard BD theory as well as a few particular
types of the scalar tensor theories. It is clear that such a task can be done only
by considering the two sets of the MBDT field equations which
correspond to those derived from the standard BD action including a scalar potential.
As mentioned, there are two other sets which have not any analog in the standard
BD theory. How can we interpret these field equations? Under which
conditions, these interpretations contact with those obtained in IMT?

\item We should emphasize that the solutions of this manuscript are more extended
than their corresponding ones obtained in the standard BD theory as well
as those obtained in \cite{RFM14}. Moreover, a few classes of our solutions
 have no analog in comparison with those obtained by means of corresponding conventional setting.
 We should note that herein solutions still can be generalized by assuming an
 ordinary matter in the bulk and/or supposing that the BD scalar field and the components of the metric
 to be functions of the extra coordinate $l$.

\item In order to obtain the herein solutions, we have not introduced conformal time and other variables
 different from the scale factor and the BD scalar field, see, e.g.,~\cite{MW95,TV96, O97,Faraoni.book}.


\item As our MBDT setting \cite{RFM14} has been formulated in arbitrary
dimensions, thus, it can be employed not only for obtaining the reduced
solutions (for a specified model) in $(3+1)$-dimensions but also it
can be examined for deriving the solutions in $(2+1)$-dimensions.
This manner of studying lower dimensional gravity theories (see, e.g.,~\cite{RRT95} and references therein)
can be of interest in probing a relationship between such theories
and the standard  four dimensional standard BD theory, as well as introducing
 a procedure of producing exact solutions in $(2+1)$
dimensions that are, as might be expected, related to the vacuum $(3+1)$-dimensional solutions.
Subsequently, we can also investigate what happen for the behaviors of
the physical quantities in the particular cases,
especially, when the BD coupling parameter goes to infinity.
\end{itemize}
\section*{Acknowledgments}
SMMR appreciates for the support of grant SFRH/BPD/82479/2011 by
the Portuguese Agency Funda\c{c}\~ao para a Ci\^encia e
Tecnologia. This research
is supported by the grants CERN/FP/123618/2011 and UID/MAT/00212/2013.



\begin{thebibliography}{1}
\bibitem{RFM14}S.M. M. Rasouli, M. Farhoudi and P. V. Moniz, Classical Quantum Gravity \textbf{31}, 115002 (2014).
\bibitem{Faraoni.book}V. Faraoni, \textit{Cosmology in Scalar Tensor Gravity}     (Dordrecht:Kluwer Academic, 2004).
\bibitem{D37}P. A. M., Dirac, \textit{Nature} \textbf{139}, 323 (1937).
\bibitem{D38}P. A. M., Dirac, \textit{Proc. Roy. Soc. (London)} \textbf{A165}, 199 (1938).
\bibitem{J66}P. Jordan, \textit{Die Expansion der Erde} Friedrich Vieweg and Sohn, Braunschweig.
\bibitem{J55}P. Jordan, \textit{Schwerkraft and Weltall} Friedrich Vieweg and Sohn, Braunschweig, p. 128.
\bibitem{J48}P. Jordan, \textit{Astron. Nachr} \textbf{276}, 193 (1948).
\bibitem{T48} Y. Thiry, \textit{Compt. Rend. Acad. Sci. (Paris)} \textbf{226}, 216 (1948).
\bibitem{BD61}C. Brans and R.H. Dicke, {\it Phys. Rev.} \textbf{124}, 925 (1961).
\bibitem{o'hanlon-tupper-72}J. O'Hanlon and B.O.J. Tupper, {\it Nuovo Cim.} \textbf{7B}, 305 (1972).
\bibitem{DO71} H. Dehnen and O. Obreg$\acute{\rm o}$n, \textit{Astrophys. Space Sci.} \textbf{14}, 454 (1971).
\bibitem{DO72a} H. Dehnen and O. Obreg$\acute{\rm o}$n, \textit{Astrophys. Space Sci.} \textbf{15}, 326 (1972).
\bibitem{DO72b} H. Dehnen and O. Obreg$\acute{\rm o}$n, \textit{Astrophys. Space Sci.} \textbf{17}, 338 (1972).
\bibitem{M78}A. Miyazaki, {\it Phys. Rev. Lett.} \textbf{40}, 11 (1978).
\bibitem{C83} P. Chauvet, \textit{Astrophys. Space Sci.} \textbf{90}, 51 (1983).
\bibitem{CN91} P. Chauvet and H. N. $N\acute{U}\tilde{N}$EZ-Y$\acute{E}$PEZ, \textit{Astrophys. Space Sci.} \textbf{178}, 165 (1991).
\bibitem{BP97}J. D. Barrow and P. Parsons, \textit{Phys. Rev. D} \textbf{55}, 1906 (1997).
\bibitem{LP-rev} D. Lorenz-Petzold, \textit{Astrophys. Space Sci.} \textbf{98}, 101 (1984).
\bibitem{qiang2005}L. Qiang, Y. Ma, M. Han and D. Yu, {\it Phys. Rev. D} \textbf{71}, 061501 (2005).
\bibitem{qiang2009}Li-e Qiang, Yan Gonga, Yongge Ma and Xuelei Chena, {\it Phys. Lett. B} \textbf{681}, 210 (2009).
\bibitem{RFK11}S. M. M. Rasouli, M. Farhoudi and N. Khosravi, \textit{Gen. Rel. Grav.} \textbf{43}, 2895 (2011).
\bibitem{LF15}N. A. Lima, P. G. Ferreira, \textit{On the phenomenology of extended Brans-Dicke Gravity}, arXiv:1506.07771 [astro-ph.CO].
\bibitem{PW92}P.S. Wesson and J. Ponce de Leon, {\it J. Math. Phys.} \textbf{33}, 3883 (1992).
\bibitem{stm99}P.S. Wesson, \textit{Space--Time--Matter: Modern Kaluza--Klein Theory} (World Scientific, Singapore, 1999).
\bibitem{DRJ09} N. Doroud, S.M. M. Rasouli and S. Jalalzadeh,
               {\it Gen. Rel. Grav.} \textbf{41}, 2637 (2009).
\bibitem{RJ10} S.M. M. Rasouli and S. Jalalzadeh,
              {\it Ann. Phys. (Berlin)} \textbf{19}, 276 (2010).
\bibitem{Ponce1}J. Ponce de Leon, {\it Class. Quant. Grav.} \textbf{27}, 095002 (2010).
\bibitem{Ponce2}J. Ponce de Leon, {\it JCAP} \textbf{03}, 030 (2010).
\bibitem{RFS11}S.M. M. Rasouli, M. Farhoudi and H.R. Sepangi, {\it Class. Quant. Grav.} \textbf{28}, 155004 (2011).
\bibitem{BP01}N. Banerjee and D. Pavon, \textit{Phys. Rev. D} \textbf{63}, 043504 (2001).
\bibitem{SS01} S. Sen and A.A. Sen, \textit{Phys. Rev. D} \textbf{63}, 124006 (2001).
\bibitem{KS91}V. A. Kosteleckji  and  S. Samuel, {\it Phys.  Lett.  B} \textbf{270}, 21 (1991).
\bibitem{OW97}J.M. Overduin and P.S. Wesson, {\it Phys. Rep.} \textbf{283}, 303 (1997).
\bibitem{M86}K. Maeda, {\it Phys. Letters} \textbf{166B}, 59 (1986).
\bibitem{W06}C. M. Will, \textit{Living Rev. Rel.} \textbf{9}, 3 (2006).
\bibitem{N68}K. Nordtvedt, \textit{Phys. Rev. D} \textbf{169}, 1017 (1968).
\bibitem{EP01}G. Esposito-Farese and D. Polarski, {\it Phys. Rev. D} \textbf{63}, 063504 (2001).
\bibitem{NP07} S. Nesseris and L. Perivolaropoulos, {\it Phys. Rev. D} \textbf{75}, 023517 (2007).
\bibitem{B93}J. D. Barrow, \textit{Phys. Rev. D} \textbf{47}, 5329 (1993).
\bibitem{RM14} S. M. M. Rasouli and P. V. Moniz, Phys. Rev. D \textbf{90}, 083533 (2014).
\bibitem{LP83a} D. Lorenz-Petzold, \textit{Astrophys. Space Sci.} \textbf{96}, 451 (1983).
\bibitem{CE83} J. M. Cerver\'{o} and P. G. Est\'{e}vez, \textit{Gen. Rel. Grav.} \textbf{15}, 351 (1983).
\bibitem{Planck2013} Planck 2013 results. XVI. Cosmological parameters, arXiv:1303.5076.
\bibitem{M82} A. Miyazaki, \textit{Nuovo Cimento} \textbf{68B}, 126 (1982).
\bibitem{M01} A. Miyazaki, Arxive: gr-qc/0012104.
\bibitem{MW95}J.P. Mimoso and D. Wands {\it Phys. Rev. D} \textbf{51}, 477 (1995).
\bibitem{O97} A. Oukuiss, \textit{Nucl. Phys. B} \textbf{486}, 413 (1997).
\bibitem{TV96} D. F. Torres and H. Vucetich, \textit{Phys. Rev. D} \textbf{54}, 7373  (1996).
\bibitem{RF76}V. A. Ruban and A. M. Finkelstein, {\it Astrofizika} \textbf{12}, 371 (1976).
\bibitem{SD89}R. K. Tarachand Singh and A. Ratnaprabha, {\it Gen. Rel. Grav.} \textbf{21}, 1249 (1989).
\bibitem{XY93}H. Xing and W. You-lin, {\it Gen. Rel. Grav.} \textbf{25}, 1 (1993).
\bibitem{Pon01}J. Ponce de Leon, {\it Mod. Phys. Lett. A} \textbf{16}, 2291 (2001).
\bibitem{Giostri12} R. Giostri, M. Vargas dos Santos, I. Waga, R. R. R. Reis, M. O. Calv$\tilde{a}$o, B. L. Lago, 2012, \textit{JCAP} \textbf{03}, 027 (2012).
\bibitem{RRT95}S. Rippl, C. Romero and R. Tavakol, {\it Class. Quant. Grav.} \textbf{12}, 2411 (1995).






































\end{thebibliography}
\end{document}